\pdfoutput=1
\documentclass[acmsmall,screen,nonacm]{acmart}

\usepackage{mathpartir}
\usepackage{tikz-cd}
\usepackage{enumitem}
\usepackage{wrapfig}
\usepackage{fancyvrb}
\usepackage{xspace}
\usepackage{thmtools} %
\usepackage{mathtools}
\usepackage{hyperref}
\usepackage[capitalise]{cleveref}
\usepackage[dvipsnames]{xcolor}

\usepackage[LGR,T1]{fontenc}
\DeclareMathAlphabet{\mathgtt}{LGR}{cmtt}{m}{n}

\colorlet{keywordcolor}{blue}
\colorlet{nonlintycolor}{RubineRed}
\colorlet{lintycolor}{PineGreen}
\colorlet{nonlinconstructorcolor}{nonlintycolor}
\colorlet{linconstructorcolor}{lintycolor}
\colorlet{alphabetcolor}{black}
\colorlet{funarrowcolor}{lintycolor}
\definecolor{backcolour}{rgb}{0.95,0.95,0.92}

\usepackage{listings}
\lstdefinelanguage{Agda}{
alsoletter={0123\&'},
keywords={data, where, module, import, open, public,
record, field, let, in, if, then, else, case, of, with,
do, postulate, primitive, mutual, abstract, private,
forall, exists, cong, set, prop, sort, Level, Data, Type,
Renamer},
morekeywords=[2]{L, Bool, Nat, Fin},
morekeywords=[3]{String, Dyck, Trace, Exp, Atom, O, D, C, A, NotStartsWithLP, NotStartsWithRP},
morekeywords=[4]{bal, .nil, .cons, cons, nil, inl, inr, stop, 1to1, 1to2, 0to2, 0to1,
  done, add, num, parens, left, unexpected, lookAheadRP, lookAheadNot, closeGood, closeBad,
  doneGood, doneBad},
morekeywords=[5]{.states, states, .isAcc, isAcc, transitions, src, dst, .transitions, .dst,
  .src, .label, label, 0 , 1 , 2, 3, true, false},
morekeywords=[6]{NUM},
morekeywords=[7]{\&, '},
sensitive=true,
comment=[l]{--},
morecomment=[s]{\{-}{-\}},
morestring=[b]",
mathescape=true,
moredelim={[is][]{\#}{\#}},
escapeinside={(*@}{@*)},
escapebegin=\color{funarrowcolor},
escapeend={}
}
\usepackage{placeins} %
\usepackage{array}     %

\lstnewenvironment{floatlisting}{
  \FloatBarrier
  \lstset{
    aboveskip=2mm
  }
}{}

\usepackage{stmaryrd}
\usepackage{scalerel}
\usepackage{tikz}

\usepackage{pdfpages}

\newif\ifdraft
\draftfalse

\newcommand{\Cl}{\mathsf{Cl}}

\newcommand{\id}{\mathsf{id}}

\newcommand{\define}[2]{#1 \coloneqq #2}
\newcommand{\prn}[1]{(#1)}
\newcommand{\cat}[1]{\mathcal{#1}}
\newcommand{\inob}[2]{#1 : \mathcal{#2}}
\newcommand{\optext}{\mathsf{op}}
\newcommand{\opsuperscript}[1]{#1^{\optext}}
\newcommand{\op}[1]{\opsuperscript{\cat{#1}}}
\newcommand{\seqstar}[2]{#1 \mathbin{\star} #2}

\newcommand{\seq}[2]{\seqstar{#1}{#2}}
\newcommand{\cathom}[3]{\cat{#1}\prn{#2,#3}}
\newcommand{\cathomraw}[3]{{#1}\prn{#2,#3}}
\newcommand{\incathom}[4]{#1 : \cathom{#2}{#3}{#4}}
\newcommand{\incathomraw}[4]{#1 : \cathomraw{#2}{#3}{#4}}
\newcommand{\dispsuperscript}[1]{\overline{#1}}
\newcommand{\vertsuperscript}[1]{\overline{#1}^{\mathbf{V}}}
\newcommand{\mkdisp}[1]{\dispsuperscript{#1}}
\newcommand{\disp}[1]{\dispsuperscript{\mathcal{#1}}}
\newcommand{\disppsh}[1]{\dispsuperscript{#1}}
\newcommand{\vertpsh}[1]{\vertsuperscript{#1}}
\newcommand{\catofelts}[1]{\mathsf{Elt}\prn{#1}}
\newcommand{\dispob}[2]{\disp{#1}_{#2}}

\newcommand{\indispob}[3]{#1 : \dispob{#2}{#3}}
\newcommand{\disphom}[4]{\disp{#1}_{#2}\prn{#3,#4}}

\newcommand{\indisphom}[5]{#1 : \disphom{#2}{#3}{#4}{#5}}
\newcommand{\dispopsuperscript}[1]{{\prn{#1}}^{\optext}}
\newcommand{\dispop}[1]{\dispopsuperscript{\disp{#1}}}
\newcommand{\fibercat}[2]{{\disp{#1}}_{#2}}
\newcommand{\fiberofexpr}[2]{{\prn{#1}}_{#2}}
\newcommand{\reindex}[2]{\mathrm{reindex}_{#1}\prn{#2}}
\newcommand{\totalcat}[1]{\textstyle\int #1}

\newcommand{\termcat}{\mathbf{1}}
\newcommand{\Set}{\mathsf{Set}}
\newcommand{\SetD}{\dispsuperscript{\Set}}
\newcommand{\singletonset}{\{\ast\}}
\newcommand{\Fam}[1]{\mathsf{Fam}\prn{#1}}
\newcommand{\Psh}{\mathsf{Psh}}
\newcommand{\wPsh}{\omega\Psh}
\newcommand{\FamwPsh}{\Fam{\wPsh}}
\newcommand{\PshD}{\dispsuperscript{\Psh}}
\newcommand{\Prop}{\mathsf{Prop}}
\newcommand{\PropD}{\dispsuperscript{\mathsf{Prop}}}
\newcommand{\flag}[1]{\texttt{-{}-#1}}
\newcommand{\Type}{\mathsf{Type}}

\newcommand{\elt}[2]{\prn{#1, #2}}
\newcommand{\pshelt}[2]{#1~#2}
\newcommand{\dispelt}[2]{\prn{#1, #2}}
\newcommand{\disppshelt}[3]{\disppsh{#1}\prn{#2, #3}}

\newcommand{\PiTy}[3]{\textstyle\prod_{#1 : #2} #3}
\newcommand{\SigTy}[3]{\textstyle\sum_{#1 : #2} #3}

\newcommand{\Mon}[1]{\mathrm{Mon}[#1]}
\newcommand{\CCCtxt}{\mathrm{CCC}}
\newcommand{\CCC}[1]{\CCCtxt[#1]}
\newcommand{\BiCCC}[1]{\mathrm{BiCCC}[#1]}
\newcommand{\CCtxt}{\mathrm{CC}}
\newcommand{\CC}[1]{\CCtxt[#1]}
\newcommand{\lang}{\mathcal{L}}
\newcommand{\langsig}[1]{\lang [#1]}

\newcommand{\dom}{\mathsf{dom}}
\newcommand{\cod}{\mathsf{cod}}
\newcommand{\rec}{\mathsf{rec}}
\newcommand{\recinterp}[1]{\rec_{#1}}

\newcommand{\recinterpCCC}[1]{\rec^{\CCCtxt}_{#1}}
\newcommand{\elim}{\mathsf{elim}}
\newcommand{\eliminterp}[1]{\elim_{#1}}
\newcommand{\dispiota}{\mkdisp{\iota}}
\newcommand{\bool}{\mathsf{Bool}}
\newcommand{\boolt}{\mathsf{true}}
\newcommand{\boolf}{\mathsf{false}}
\newcommand{\Canonical}{\mathsf{Canon}}
\newcommand{\tmtt}{\mathsf{tt}}
\newcommand{\tmff}{\mathsf{ff}}
\newcommand{\tmz}{\mathsf{z}}
\newcommand{\tms}{\mathsf{s}}
\newcommand{\suc}{\mathsf{suc}}
\newcommand{\QuoteBool}{\mathrm{QuoteBool}}
\newcommand{\QuoteNat}{\mathrm{QuoteNat}}
\newcommand{\IdFun}{\mathrm{Id}}
\newcommand{\reindop}{\mathsf{subst}}
\newcommand{\List}{\mathsf{List}}
\newcommand{\nil}{\mathsf{nil}}
\newcommand{\cons}{\mathsf{cons}}
\newcommand{\ListOf}[1]{\List~#1}
\newcommand{\reclist}{\rec_{\List}}
\newcommand{\elimlist}{\elim_{\List}}

\usepackage{CJKutf8}
\newcommand{\yochar}{\text{\begin{CJK}{UTF8}{min}よ\end{CJK}}}
\newcommand{\yo}[1]{\yochar\,#1}
\newcommand{\nervetxt}{N}
\newcommand{\nerve}[1]{\nervetxt_{#1}}
\newcommand{\dispyochar}{\dispsuperscript{\yochar}}
\newcommand{\vertyochar}{\vertsuperscript{\yochar}}
\newcommand{\dispyo}[1]{\dispyochar\,#1}
\newcommand{\recyo}[1]{\rec_{\yochar}\prn{#1}}
\newcommand{\recdispyo}[1]{\rec_{\dispyochar}\prn{#1}}
\newcommand{\recvertyo}[1]{\rec_{\vertyochar}\prn{#1}}
\newcommand{\IsoFiber}{\mathrm{IsoFiber}}
\newcommand{\IsoComma}{\mathrm{IsoComma}}
\newcommand{\Iso}{\mathrm{Iso}}

\newcommand{\cartlift}[2]{#1^{*} #2}
\newcommand{\cartliftbare}[1]{#1^{*}}
\newcommand{\vertprod}[2]{#1 \times^{\mathbf{V}} #2}
\newcommand{\dispprod}[2]{#1 \mathbin{\overline{\times}} #2}
\newcommand{\vertexp}[2]{#1 \Rightarrow^{\mathbf{V}} #2}
\newcommand{\dispexp}[2]{#1 \mathrel{\overline{\Rightarrow}} #2}
\newcommand{\dispcoprod}[2]{#1 \mathbin{\overline{+}} #2}
\newcommand{\dispterm}{\overline{1}}
\newcommand{\dispinit}{\overline{0}}
\newcommand{\uq}[2]{\forall_{#1}\,#2}
\newcommand{\later}{\mathord{\triangleright}}
\newcommand{\vertlater}{\later^{\mathbf{V}}}
\newcommand{\nextmap}{\mathsf{next}}
\newcommand{\lamforall}[1]{\lambda^{\forall}_{#1}}
\newcommand{\appforall}[1]{\mathsf{app}^{\forall}_{#1}}

\newcommand{\app}{\mathsf{app}}

\newcommand{\pext}[1]{\mathrm{ext}\prn{#1}}
\newcommand{\pshtens}[2]{#1 \otimes #2}
\newcommand{\larghom}[2]{#1 \Rightarrow #2}
\newcommand{\sqprn}[1]{\left[ #1 \right]}

\newcommand{\slicecat}[2]{#1 / #2}

\newcommand{\nat}{\mathsf{Nat}}

\newcommand{\inl}{\mathsf{inl}}
\newcommand{\inr}{\mathsf{inr}}

\newcommand{\emptyty}{0}
\newcommand{\unitty}{1}
\newcommand{\unitel}{\mathsf{tt}}

\newcommand{\sem}[1]{\llbracket #1 \rrbracket}

\newcommand{\interpIn}[2]{#1 \leadsto #2}

\newcommand{\absty}{X}
\newcommand{\SigFlip}{\Sigma_{\mathsf{flip}}}
\newcommand{\opinit}{\mathsf{init}}
\newcommand{\opflip}{\mathsf{flip}}
\newcommand{\opread}{\mathsf{read}}
\newcommand{\semwith}[2]{\sem{#2}_{#1}}
\newcommand{\semB}[1]{\semwith{\bool}{#1}}
\newcommand{\semN}[1]{\semwith{\nat}{#1}}
\newcommand{\notb}{\mathsf{not}}
\newcommand{\even}{\mathsf{even}}
\newcommand{\lrel}[1]{R_{#1}}
\newcommand{\lrelat}[3]{\lrel{#1}\prn{#2, #3}}
\newcommand{\Agda}{\raisebox{-0.2ex}{\includegraphics[height=1.8ex]{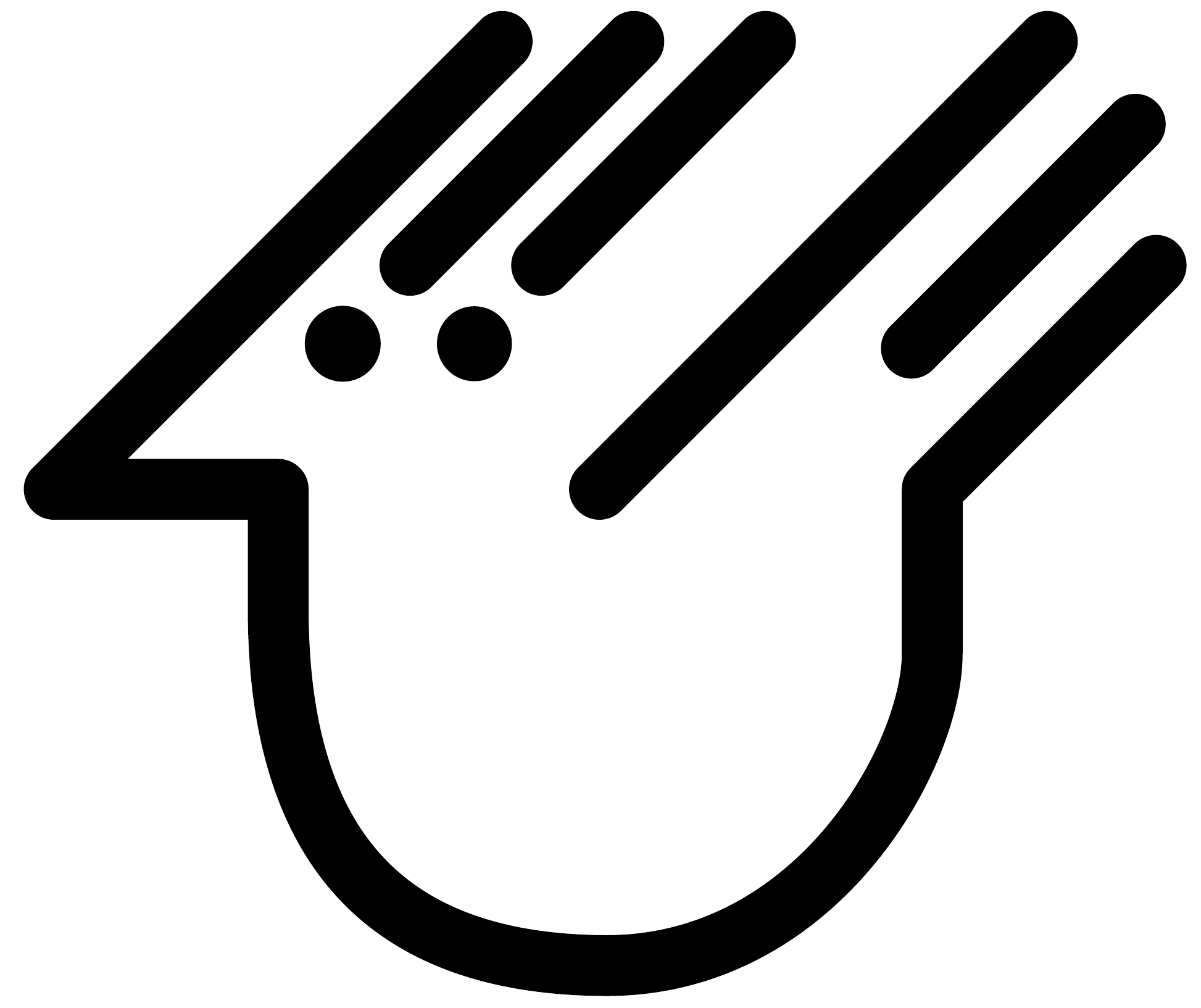}}\xspace}

\newif\ifwithappendix
\ifdefined\noappendix\withappendixfalse\else\withappendixtrue\fi

\newif\ifagdaexternal \agdaexternaltrue

\newcommand{\agdaentry}[4]{%
  \expandafter\def\csname agdarepo@#1\endcsname{#2}%
  \expandafter\def\csname agdapath@#1\endcsname{#3}%
  \expandafter\def\csname agdaline@#1\endcsname{#4}}
\newcommand{\AgdaExtURL}[1]{%
  \csname agdabase\csname agdarepo@#1\endcsname\endcsname
  \csname agdapath@#1\endcsname\#L\csname agdaline@#1\endcsname}
\DeclareRobustCommand{\AgdaRef}[1]{%
  \ifagdaexternal \href{\AgdaExtURL{#1}}{\Agda}%
  \else \ifwithappendix \hyperlink{agda:#1}{\Agda}\else \Agda \fi\fi}
\agdaentry{def:dispcat}{up}{Cubical/Categories/Displayed/Base.agda}{20}
\agdaentry{def:setd}{safe}{Cubical/Categories/Displayed/Instances/Sets/Base.agda}{31}
\agdaentry{def:section}{up}{Cubical/Categories/Displayed/Section/Base.agda}{152}
\agdaentry{thm:elim}{safe}{Cubical/Categories/Instances/Free/BiCartesianClosedCategory/Forded.agda}{258}
\agdaentry{thm:canonicity}{safe}{Gluing/BiCartesianClosedCategory/BoolNatCanonicity/Forded.agda}{124-L128}
\agdaentry{lambda:free}{safe}{Cubical/Categories/Instances/Free/Lambda/Negative.agda}{79}
\agdaentry{lambda:canonicity}{safe}{Gluing/Lambda/BoolNatCanonicity/Negative.agda}{87}
\agdaentry{def:totalcat}{up}{Cubical/Categories/Instances/TotalCategory/Base.agda}{29}
\agdaentry{def:reindex}{up}{Cubical/Categories/Displayed/Instances/Reindex/Base.agda}{90}
\agdaentry{def:catofelts}{safe}{Cubical/Categories/Displayed/Instances/Graph/Presheaf.agda}{61}
\agdaentry{def:slicecat}{safe}{Cubical/Categories/Displayed/Presheaf/Uncurried/Base.agda}{95}
\agdaentry{def:displayedpsh}{safe}{Cubical/Categories/Displayed/Presheaf/Uncurried/Base.agda}{203}
\agdaentry{def:dispyo}{safe}{Cubical/Categories/Displayed/Presheaf/Uncurried/Representable.agda}{64}
\agdaentry{def:reindexdisppsh}{safe}{Cubical/Categories/Displayed/Presheaf/Uncurried/Base.agda}{314}
\agdaentry{def:disppshmor}{safe}{Cubical/Categories/Displayed/Presheaf/Uncurried/Base.agda}{466}
\agdaentry{def:dispunivelt}{safe}{Cubical/Categories/Displayed/Presheaf/Uncurried/Representable.agda}{168}
\agdaentry{def:verticalpsh}{safe}{Cubical/Categories/Displayed/Presheaf/Uncurried/Base.agda}{212}
\agdaentry{def:vertunivelt}{safe}{Cubical/Categories/Displayed/Presheaf/Uncurried/Representable.agda}{141}
\agdaentry{def:vertterm}{safe}{Cubical/Categories/Displayed/Presheaf/Uncurried/UniversalProperties.agda}{63}
\agdaentry{def:vertprod}{safe}{Cubical/Categories/Displayed/Presheaf/Uncurried/UniversalProperties.agda}{81}
\agdaentry{def:vertbp}{safe}{Cubical/Categories/Displayed/Presheaf/Uncurried/UniversalProperties.agda}{87}
\agdaentry{thm:setd-fibration}{safe}{Cubical/Categories/Displayed/Instances/Sets/Properties.agda}{93}
\agdaentry{def:vertexp}{safe}{Cubical/Categories/Displayed/Presheaf/Uncurried/Constructions/Exponential.agda}{379}
\agdaentry{def:uq}{safe}{Cubical/Categories/Displayed/Presheaf/Uncurried/Constructions/UniversalQuantifier.agda}{189}
\agdaentry{def:vertcc}{safe}{Cubical/Categories/Displayed/Limits/CartesianClosedV.agda}{38}
\agdaentry{thm:setd-vertical}{safe}{Cubical/Categories/Displayed/Instances/Sets/Properties.agda}{204}
\agdaentry{thm:vert-cart}{safe}{Cubical/Categories/Displayed/Limits/CartesianV'.agda}{53}
\agdaentry{thm:vert-cc}{safe}{Cubical/Categories/Displayed/Limits/CartesianClosedV.agda}{66}
\agdaentry{cor:setd-dispcc}{safe}{Cubical/Categories/Displayed/Instances/Sets/Properties.agda}{209}
\agdaentry{def:predd}{safe}{Cubical/Categories/Displayed/Instances/Family/Base.agda}{20}
\agdaentry{def:pshd}{safe}{Cubical/Categories/Displayed/Instances/Presheaf/Eq/Base.agda}{273}
\agdaentry{thm:pshd-vertcc}{safe}{Cubical/Categories/Displayed/Instances/Presheaf/Eq/CartesianClosed.agda}{155}
\agdaentry{thm:cocont-adjoint}{safe}{Cubical/Categories/Presheaf/StrictHom/RightAdjoint.agda}{107}
\agdaentry{def:isoD}{safe}{Cubical/Categories/Displayed/Instances/Arrow/Base.agda}{61}
\agdaentry{def:isofiber}{safe}{Cubical/Categories/Displayed/Instances/IsoFiber/Base.agda}{43}
\agdaentry{thm:isofiber-reflection}{safe}{Cubical/Categories/Displayed/Instances/IsoFiber/Base.agda}{47}
\agdaentry{def:isocomma}{safe}{Cubical/Categories/Displayed/Instances/Arrow/Monoidal.agda}{86}
\agdaentry{thm:isocomma-reflection}{safe}{Cubical/Categories/Displayed/Instances/Arrow/Base.agda}{85}
\agdaentry{thm:conservativity}{safe}{Gluing/CartesianClosedCategory/Conservativity/OverCartesianCategories/Forded.agda}{238}
\agdaentry{lem:thin-isoretract}{safe}{Cubical/Categories/HLevels.agda}{52}
\agdaentry{thm:coherence}{safe}{Cubical/Categories/Instances/Free/Monoidal/Coherence.agda}{29}
\agdaentry{thm:stepindexed}{safe}{Gluing/Category/GuardedFixedPoint.agda}{257}
\agdaentry{synthetic}{unsafe}{Guarded/Gluing/Canonicity.agda}{210}
\agdaentry{frobenius}{safe}{Cubical/Categories/Displayed/Presheaf/Uncurried/Constructions.agda}{338}
\agdaentry{beck-chevalley}{safe}{Cubical/Categories/Displayed/Presheaf/Uncurried/Constructions.agda}{410}
\agdaentry{lem:identity-extension}{safe}{Gluing/BiCartesianClosedCategory/IdentityExtension.agda}{85}
\agdaentry{thm:repind}{safe}{Gluing/BiCartesianClosedCategory/FlipperParametricity.agda}{149}
\agdaentry{thm:parametricity-stack}{safe}{Gluing/BiCartesianClosedCategory/StackParametricity.agda}{181}

\declaretheoremstyle[headfont=\normalfont\itshape]{defstyle}
\declaretheoremstyle[headfont=\normalfont\scshape]{thmstyle}

\declaretheorem[name=Definition, style=defstyle, numberwithin=section]{definition}
\declaretheorem[style=thmstyle, name=Theorem, numberlike=definition]{theorem}
\declaretheorem[style=thmstyle, name=Lemma, numberlike=definition]{lemma}
\declaretheorem[style=thmstyle, name=Corollary, numberlike=definition]{corollary}
\declaretheorem[style=thmstyle, name=Construction, numberlike=definition]{construction}
\declaretheorem[style=defstyle, name=Intensional Remark, numberlike=definition]{intensionalremark}

\crefname{definition}{Definition}{Definitions}
\crefname{theorem}{Theorem}{Theorems}
\crefname{corollary}{Corollary}{Corollaries}
\crefname{construction}{Construction}{Constructions}
\crefname{intensionalremark}{Intensional Remark}{Intensional Remarks}

\begin{document}

\pagebreak

\title{Logical Relations on Display}

\author{Steven Schaefer}
\orcid{0009-0007-1258-9501}
\affiliation{%
  \institution{University of Michigan}
  \city{Ann Arbor}
  \country{USA}
}
\email{stschaef@umich.edu}

\author{Johnson He}
\orcid{0009-0004-3358-0643}
\affiliation{%
  \institution{Indiana University}
  \city{Bloomington}
  \country{USA}
}
\additionalaffiliation{%
  \institution{University of Michigan}
  \city{Ann Arbor}
  \country{USA}
}
\email{hejo@iu.edu}

\author{Max S. New}
\orcid{0000-0001-8141-195X}
\affiliation{%
  \institution{University of Michigan}
  \city{Ann Arbor}
  \country{USA}
}
\email{maxsnew@umich.edu}

\begin{abstract}
  The metatheory of programming languages suffers from a lack of
  reuse. Researchers develop individual calculi and prove standard
  properties, but very few results are truly reused across papers.
  The problem is worse in proof assistants, where developments
  routinely duplicate code from prior work or rebuild it from scratch.

  Category theory provides the most promising path toward reusable
  programming language semantics. Universal properties capture the
  type connectives---products, functions, sums---together with their
  $\beta$/$\eta$ laws. Free categorical structures identify syntactic
  calculi. \emph{Displayed} categories provide a uniform language for
  specifying logical relations, and general theorems about fibrations
  allow their structure to be assembled compositionally. In this work
  we develop a library of formalized category theory in the Cubical
  Agda proof assistant directly applicable to programming language
  metatheory. We develop theories of universal properties and free
  categorical structures, apply them to define denotational semantics
  and logical relations for simply typed calculi, and identify
  encodings of the categorical constructions that make typechecking
  them tractable. From the same categorical components, we
  construct several familiar results via a logical relations construction---canonicity,
  parametricity,
  conservativity---and provide a proof of coherence for monoidal categories.
\end{abstract}

\maketitle
\section{Introduction}
\label{sec:intro}

Programming language theory is, in large part, the study of
metatheoretic properties of formal systems. A large class
of metatheoretic proofs proceeds by constructing a semantic model
whose structure encodes the property of interest. The syntax of the
language is interpreted in the model, and the desired conclusion is
read off from the interpretation. The model does the heavy lifting:
showing that a carefully chosen semantic domain carries
the algebraic structure of the language exposes the essence of a
metatheoretic proof. Logical relations proofs are the archetype of
this pattern: the model interprets each type as a relation on (or a
family over) the terms of the language, and the metatheoretic
property follows from the existence of the
interpretation~\cite{tait1967, plotkin1973}. Each language determines the right notion of
model. The choice of language and semantic domain varies across
applications, while the architecture of the proof does not.

Proof assistants are an attractive setting for metatheoretic proofs
of this form. They provide machine-checked guarantees of correctness,
eliminating the risk of subtle errors in complex reasoning. However, proof
assistants demand total precision from their users. Faced with this
requirement, proof developers tend toward low-level, concrete
descriptions. These direct encodings of the specific argument at hand
resist reuse and create a culture of ad hoc, single-use formal
artifacts.

The lack of reuse is especially unfortunate given the regularity of
the arguments involved. Metatheoretic proofs share a common architecture---construct a
model, interpret the syntax, extract the conclusion---yet there is no
standard repository of theorems and constructions from which to
assemble such proofs. Each formalization effort is
labor-intensive, and the elegant mathematical structure that can
organize these proofs on paper is lost when artifacts are recreated
from scratch at the lowest level of abstraction in each new project.
The outcome is massive duplication across projects that
are, at a high level, doing the same thing.

\looseness=-1 Category theory provides a natural framework for this reuse. Its abstractions---universal properties, functoriality,
adjunctions---organize metatheoretic arguments at a higher level of
generality. On paper,
category-theoretic treatments of programming language semantics are
well established~\cite{lambek1986higher, crole1994categories,
JacobsCLTT}, yet proof
developers have been slow to adopt this perspective in practice.
Formalized libraries of category theory
exist~\cite{1lab, UniMath, hu2021formalizing, mathlib2020},
but they are organized as general mathematics rather than around the
semantic constructions that metatheoretic proofs assemble. We work
backwards from the application, providing exactly the infrastructure
that programming language metatheorems demand.

Category theory itself poses
challenges when formalized in intensional type theory. Classical
presentations of key concepts often require reasoning about equality
of objects in a category, a condition that is trivial on paper but
becomes a serious obstacle in a formal
setting~\cite{ahrens2019displayed}. For
instance, the standard definition of a fibration asks for lifts that
are equal to a given object on the nose, introducing equational
constraints which propagate through constructions that depend on
it.

\looseness=-1 These equational constraints also carry a performance cost. A
proliferation of equality obligations is tedious for the user to
manage and expensive for the typechecker to resolve. The wrong
presentation of the underlying mathematics can make formalization
prohibitively slow, turning a routine verification into
an intractable one.

In this paper, we develop a library of reusable categorical
abstractions for mechanizing metatheory in Cubical
Agda~\cite{VezzosiMortbergAbel2019}.
We make extensive use of \emph{displayed
categories}~\cite{ahrens2019displayed}, a generalization of category
theory that more naturally applies to dependently typed
constructions.

More broadly, we advocate a methodology: formal proof developers should
build metatheoretic arguments with modular categorical
components rather than starting from scratch.
Each type connective of a programming language---products, function
types, and so on---is determined by a \emph{universal property}.
Universal properties describe the algebraic interface of each type
former, dictating how it must be interpreted in any semantics; and
because every connective is an instance of a universal property,
theorems about universal properties apply to all connectives at once,
rather than being reproven for each.
Our contributions realize this reuse, and are as follows:
\begin{enumerate}
  \item A compositional framework for mechanized metatheory
    (\cref{sec:free-cats,sec:dispuniv}). The syntax of a programming
    language is a free category with a dependent elimination principle:
    to eliminate out of the syntax, one constructs a displayed category
    over it with the appropriate universal structures. We present these
    structures as universal elements of displayed presheaves, in a
    displayed and a vertical form, with promotion theorems that turn
    fiberwise vertical structure into the displayed structure the
    eliminator consumes.
  \item A library of reusable semantic domains in Cubical Agda
    (\cref{sec:concrete,sec:implementation})---displayed categories of
    sets, propositions, presheaves, and isomorphisms.
  \item Mechanized proofs of canonicity, parametricity,
    conservativity, coherence for monoidal categories, and
    step-indexed canonicity (\cref{sec:applications}); each obtained by
    instantiating the framework in a few hundred lines of user-facing
    code.
\end{enumerate}

The library comprises 243 modules---roughly 35{,}000 lines---of
Cubical Agda, extending the \texttt{agda/cubical}
library~\cite{cubicalagdalib}.
It is in the \flag{safe} fragment of Agda, except for an
alternative synthetic development of the step-indexed example
(\cref{sec:stepindexed}) that necessitates the guarded extension to Cubical Agda.
When writing a metatheoretic proof, the user appeals to the
library's structures directly rather than re-establishing them, and
spends their effort on the interesting part of the
argument: the interpretation of the generators of the syntax in the
chosen semantic domain.

\paragraph{Roadmap}
\cref{sec:free-cats} presents the syntax of a typed language as a free
category and states a dependent elimination principle for it as an
example. The principle is phrased with displayed categories, the
category-theoretic analogue of a dependent type: a displayed category
over $\cat{C}$ is a family of objects and morphisms lying over those
of $\cat{C}$. To map out of the syntax, one builds a displayed
category over it. We illustrate the method with a proof of
canonicity. \cref{sec:dispuniv} develops the (displayed) category
theory underlying our approach: universal elements, their displayed
and vertical generalizations, fibrations, and the promotion and
reindexing theorems, culminating in a compositional technique for assembling displayed models.
\cref{sec:concrete} showcases semantic domains that are
reused throughout our proofs and establishes some vertical structures
in them.  \cref{sec:applications} applies the same technique for proofs of
parametricity, conservativity, coherence, and
step-indexed canonicity.  \cref{sec:implementation}
presents the encodings that make the formalization tractable to
typecheck in intensional type theory.
\cref{sec:discussion} discusses related and future work.

\section{Programming Languages as Free Categories}
\label{sec:free-cats}

\looseness=-1 The type structures of a programming language correspond
to categorical \emph{universal properties}. We
give a simple language $\lang$---the simply typed $\lambda$-calculus
with unit, product, function, sum, and empty types---as an example.
Each connective
has the standard formation, introduction, and elimination rules
together with its $\beta$/$\eta$ equations---pairing and projections
for products, $\lambda$-abstraction and application for functions,
injections and case analysis for sums, and the nullary cases for
unit and empty.

A denotational semantics of $\lang$ assigns meaning to each rule: an
interpretation for each type former and term constructor, subject to
the $\beta$/$\eta$ equations.  Concretely, we could interpret types as
sets and terms as functions.  Sets are a natural starting point, but
no single semantic domain captures every property of interest. The
language shapes the domain necessary for representing it.  For
instance, modeling recursion calls for $\omega$-CPOs and continuous
maps~\cite{abramskyjung1994} or for step-indexed models in the topos
of trees, a presheaf category~\cite{birkedal2012trees}. Modeling name
binding or local state also calls for presheaf
categories~\cite{fiore1999, plotkinpower2002}. The property under
investigation also shapes the semantics: a logical relations proof
interprets types as predicates or as set-indexed families, while a
Kripke logical relation~\cite{mitchellmoggi1991}, such as in a
normalization proof~\cite{altenkirch2001nbe}, varies functorially over
a category of worlds or contexts. A formalization built around sets
and functions cannot accommodate these richer settings without being
refactored or duplicated. Category theory removes the commitment to a
single domain. Each of these domains is a category with the
appropriate universal properties, so a theorem proved for such a
category applies uniformly to several applications. As a result, a
user who initially formalizes results for a simple language can later
extend to recursion or local state while reusing their existing work,
because the categorical approach is \emph{modular}.

A category $\cat{C}$ packages the data common to any model: objects
serve as semantic types and morphisms as semantic terms. Each type
\emph{connective} of $\lang$ then corresponds to a \emph{universal
  property} that $\cat{C}$ must satisfy: the unit type to a terminal
object, product types to binary products, function types to
exponentials, sum types to binary coproducts, and the empty type to an
initial object. A category equipped with all of these is called a
\emph{bicartesian closed category}, and we adopt bicartesian closed
categories as the models of $\lang$. The bicartesian closed structure
reflects our choice of connectives. A language with a different
collection of type connectives calls for a different class of
structured categories, and the theory we develop applies to any such
choice. Throughout the paper, the language $\lang$ serves as a running
example of the general method.

\subsection{Specifying Generating Data}

To extend $\lang$ with base types and terms, we parameterize by a
signature. A \emph{signature} $\Sigma$ describes a collection of base types
$\Sigma_0$ and generating terms $\Sigma_1$. For example, we may
specify a base type of booleans with a signature $\Sigma_{\bool}$
containing a base type $\bool$ with closed terms $\boolt$ and
$\boolf$. We write $\langsig{\Sigma}$ for the extension of $\lang$
with the rules generated by $\Sigma$. Unlike the connectives of
$\lang$, the base types and terms of $\Sigma$ carry no universal
property. Interpreting them is a choice a model must make.

\begin{definition}
Given a signature $\Sigma$ and a category $\cat{C}$, an
\emph{interpretation} $\iota : \interpIn{\Sigma}{\cat{C}}$ of $\Sigma$
in $\cat{C}$ consists of a choice of object $\iota\prn{X}$ in $\cat{C}$
for each base type $X \in \Sigma_0$, and a choice of morphism
$\iota\prn{M} : \cathom{C}{\dom\prn{M}}{\cod\prn{M}}$ for each base term
$M \in \Sigma_1$.
\end{definition}

For example, an interpretation of $\Sigma_{\bool}$ in $\Set$ may
interpret $\bool$ as the set $\{0, 1\}$, with
$\iota\prn{\boolt} : \singletonset \to \{0, 1\}$ the function that
picks out $1$ and $\iota\prn{\boolf}$ the function that picks out $0$.

\subsection{The Free Category}
\label{sec:free-cats:free}

The syntax of $\langsig{\Sigma}$ itself forms a bicartesian closed
category: types are objects, terms are morphisms, and the
$\beta$/$\eta$ rules ensure the universal properties hold for each
connective. We call this category $\BiCCC{\Sigma}$, the \emph{free
bicartesian closed category over $\Sigma$}. Its terms are those of a
unary type theory, in which a context is encoded as a single type by
taking the product of its components. We work with the unary language
in the paper because it aligns more closely with the categorical
structures. The formalization also includes the negative fragment of
the $\lambda$-calculus (unit, products, and functions) with
multi-variable contexts~(\AgdaRef{lambda:free}), together with its
canonicity proof~(\AgdaRef{lambda:canonicity}). The same displayed
universal structures and gluing techniques apply to both.

\begin{figure}[b]
  \footnotesize
  \begin{alignat*}{3}
    \recinterp{\iota}\prn{A} &= \iota\prn{A}, \;\, A \in \Sigma_0
      &\qquad \recinterp{\iota}\prn{A \times B} &= \recinterp{\iota}\prn{A} \times_{\cat{C}} \recinterp{\iota}\prn{B}
      &\qquad &\reclist : \forall\, \{ B : \Type \} \to B \to \prn{A \to B \to B} \\
    \recinterp{\iota}\prn{1} &= 1_{\cat{C}}
      & \recinterp{\iota}\prn{A + B} &= \recinterp{\iota}\prn{A} +_{\cat{C}} \recinterp{\iota}\prn{B}
      & &\phantom{\reclist :{}} \to \ListOf{A} \to B \\
    \recinterp{\iota}\prn{\emptyty} &= 0_{\cat{C}}
      & \recinterp{\iota}\prn{A \Rightarrow B} &= \recinterp{\iota}\prn{A} \Rightarrow_{\cat{C}} \recinterp{\iota}\prn{B}
      & &\reclist~b~f~\nil = b \\
    & & & & &\reclist~b~f~\prn{\cons~x~xs} = f~x~\prn{\reclist~b~f~xs}
  \end{alignat*}
  \caption{The recursor of $\BiCCC{\Sigma}$ on types (left) and the
    recursor for lists (right).}
  \label{fig:rec}
\end{figure}

For any other bicartesian closed category $\cat{C}$ with an interpretation
$\iota : \interpIn{\Sigma}{\cat{C}}$ there is a unique
structure-preserving functor $\recinterp{\iota} : \BiCCC{\Sigma} \to
\cat{C}$. This functor is defined by recursion on the syntax,
mapping types to their semantic interpretation via pattern matching
(\cref{fig:rec}, left).
The action on terms is defined similarly by recursion.

$\recinterp{\iota}$ encodes the denotational semantics of
$\langsig{\Sigma}$ in $\cat{C}$: it assigns
meaning to every type and term of the language in a
structure-preserving way. It is analogous
to the recursor for an inductive data type. For instance, lists over a
type $A$ have a recursor, familiar to functional programmers as
$\mathsf{foldr}$ (\cref{fig:rec}, right).
Here $b$ interprets the empty list and $f$ interprets $\cons$. The
recursor preserves these interpretations, just as
$\recinterp{\iota}$ preserves $\langsig{\Sigma}$-model structure.

$\recinterp{\iota}$ is the \emph{only} structure-preserving functor
$\BiCCC{\Sigma} \to \cat{C}$ that agrees with $\iota$ on the
generators. This uniqueness is the initiality of $\BiCCC{\Sigma}$
among BiCCCs interpreting $\Sigma$. We
use uniqueness below to relate the recursor to the eliminator, and again in
the conservativity proof of \cref{sec:conservativity}.

\begin{definition}
\label{def:modellangsig}
Given a
bicartesian closed category
$\cat{C}$ equipped with an interpretation
$\iota : \interpIn{\Sigma}{\cat{C}}$, a \emph{model} of
$\langsig{\Sigma}$ is a structure-preserving functor
$\BiCCC{\Sigma} \to \cat{C}$.
\end{definition}

The interpretation $\iota : \interpIn{\Sigma}{\cat{C}}$ determines the
model $\recinterp{\iota}$, and by uniqueness every model arises this
way: models of $\langsig{\Sigma}$ in $\cat{C}$ correspond to
interpretations of $\Sigma$ in $\cat{C}$.

\subsection{The Elimination Principle}
\label{sec:free-cats:elim}

To prove a property of every term of a language, one argues by
\emph{induction} on the syntax, checking the property at each term
former. Metatheorems often need to construct \emph{structure} at each
term, not just verify a \emph{proposition}: a logical relation
attaches to each term a family of witnesses rather than a mere truth value. The
\emph{dependent elimination principle} covers both, letting the result
depend on the term eliminated and carry arbitrary data. In dependent
type theory, such eliminators are again familiar: for an
inductive type such as $\List$, the eliminator $\elimlist$ and its
computation rules appear in \cref{fig:elim} (right).

\begin{figure}
  \footnotesize
  \begin{alignat*}{3}
    \eliminterp{\dispiota}\prn{A} &= \dispiota\prn{A}, \;\, A \in \Sigma_0
      &\quad \eliminterp{\dispiota}\prn{A \times B} &= \dispprod{\eliminterp{\dispiota}\prn{A}}{\eliminterp{\dispiota}\prn{B}}
      &\quad &\elimlist : \forall~\{B : \ListOf{A} \to \Type\} \to B\; [\,] \\
    \eliminterp{\dispiota}\prn{1} &= \dispterm
      & \eliminterp{\dispiota}\prn{A + B} &= \dispcoprod{\eliminterp{\dispiota}\prn{A}}{\eliminterp{\dispiota}\prn{B}}
      & &\phantom{\elimlist :{}} \to \prn{\forall\, \{a\}\, \{l\} \to B\; l \to B\; \prn{\cons~a~l}} \\
    \eliminterp{\dispiota}\prn{\emptyty} &= \dispinit
      & \eliminterp{\dispiota}\prn{A \Rightarrow B} &= \dispexp{\eliminterp{\dispiota}\prn{A}}{\eliminterp{\dispiota}\prn{B}}
      & &\phantom{\elimlist :{}} \to \prn{l : \ListOf{A}} \to B \; l \\
    & & & & &\elimlist~b~f~[\,] = b \\
    & & & & &\elimlist~b~f~\prn{\cons~x~xs} = f~\prn{\elimlist~b~f~xs}
  \end{alignat*}
  \caption{The eliminator of $\BiCCC{\Sigma}$ on types (left) and the
    eliminator for lists (right).}
  \label{fig:elim}
\end{figure}

Note, the recursor and the eliminator are equally strong given the
uniqueness of the recursor. One implements $\elimlist$ using
$\reclist$ by letting the return type of the recursor be
$\SigTy{xs}{\ListOf{A}}{B~xs}$. When run on $l : \ListOf{A}$,
uniqueness guarantees that the first
projection of the result is the identity, so that the second
projection lies over the intended list. But this implementation does
not return a value at index $l$ \emph{definitionally}.
Instead, after second projection it produces a
value over a list \emph{propositionally equal} to $l$, and one must
transport over this equality to implement the interface of
$\elimlist$.  The dependent eliminator avoids this transport entirely:
$\elimlist$ is defined directly, so that the inhabitant of $B$ that it
returns is definitionally indexed by the desired list $l$.

Just as with lists, the free bicartesian closed category has a
dependent elimination principle in addition to its recursor.  The
recursor and the eliminator are again equally strong, but the eliminator is
favored because it elides equational side conditions that would arise
when simulating it with the recursor. To state the eliminator, we
must make precise the categorical analogue of a dependent type with
the concept of a \emph{displayed category}~\cite{ahrens2019displayed}.
We write composition in diagrammatic order: $\seq{f}{g}$ for $f$
followed by $g$.

\begin{definition}[{\cite[Definition 3.1]{ahrens2019displayed}}~\AgdaRef{def:dispcat}]\label{def:dispcat}
Given a category $\cat{C}$, a \emph{displayed category} $\disp{C}$ over
$\cat{C}$ assigns:
\begin{itemize}
\item to each $\inob{c}{C}$, a type $\dispob{C}{c}$ of
  \emph{displayed objects},
\item to each $\incathom{f}{C}{a}{b}$ and
  $\indispob{\mkdisp{a}}{C}{a}$, $\indispob{\mkdisp{b}}{C}{b}$, a type
  $\disphom{C}{f}{\mkdisp{a}}{\mkdisp{b}}$ of \emph{displayed morphisms}
  over $f$,
\item to each $\indispob{\mkdisp{a}}{C}{a}$, an identity
  $\id_{\mkdisp{a}} \in \disphom{C}{\id_a}{\mkdisp{a}}{\mkdisp{a}}$
  over $\id_a$, and
\item to $\mkdisp{f} \in \disphom{C}{f}{\mkdisp{a}}{\mkdisp{b}}$,
  $\mkdisp{g} \in \disphom{C}{g}{\mkdisp{b}}{\mkdisp{c}}$, a composite
  $\seq{\mkdisp{f}}{\mkdisp{g}} \in
  \disphom{C}{\seq{f}{g}}{\mkdisp{a}}{\mkdisp{c}}$ over $\seq{f}{g}$,
\end{itemize}
satisfying unitality and associativity laws lying over the
corresponding laws of $\cat{C}$.
\end{definition}

The analogy with dependent types is direct: just as $B : A \to \Type$
assigns a type to each element of $A$, a displayed category assigns a
type of displayed objects to each object of the base and a type of
displayed morphisms to each base morphism. Each displayed object lies
over a base object, just as each element of $B\,a$ lies over the point
$a$.
Our running example of a displayed category is families of sets. Its base,
$\Set$, is the category of \emph{homotopy sets}~\cite[Definition 3.1.1]{hottbook}.
\begin{definition}[\AgdaRef{def:setd}]\label{def:setd}
  The \emph{displayed category of families of sets} $\SetD$ over the category
  $\Set$
  has, over a set $X$, displayed objects the families $X \to \Set$, and
  displayed morphisms over $f : X \to Y$ between families $P$ and $Q$
  the fiberwise functions $\PiTy{x}{X}{P\prn{x} \to Q\prn{f\prn{x}}}$.
\end{definition}

\looseness=-1 Identity and composition are pointwise: the displayed identity is the
identity function, and the displayed composition of
$\mkdisp{f}$ over $f$ and $\mkdisp{g}$ over $g$ sends $x$, $p$ to
$\mkdisp{g}\prn{f\prn{x},\; \mkdisp{f}\prn{x, p}}$. All categorical
laws---left and right identity and associativity---hold
definitionally: each is proved by \texttt{refl} in Cubical Agda.

In logical-relations terms, a displayed object over a set $X$ is a
proof-relevant predicate on $X$, assigning to each element a set of
proofs. A displayed morphism over $f : X \to Y$ from $p$ to $q$ is a
proof that $f$ preserves the predicates, taking each witness for $p$
at a point $x$ to one for $q$ at $f\prn{x}$.

\begin{definition}[\AgdaRef{def:section}]\label{def:section}
  \looseness=-1 A \emph{section} of $\disp{C}$ is the categorical analogue of a
  dependent function $\PiTy{a}{A}{B\;a}$: it assigns to each base datum
  a datum lying over it. Concretely, a section chooses a displayed
  object $s\prn{c} \in \dispob{C}{c}$ for every object $c$ and a
  displayed morphism $s\prn{f} \in \disphom{C}{f}{s\prn{a}}{s\prn{b}}$
  for every morphism $\incathom{f}{C}{a}{b}$, respecting composition and
  identities.
\end{definition}

These pieces are the ingredients of a logical relation. The displayed
category $\disp{C}$ encodes the domain over which the relation is
defined; a section's object map is the relation itself, assigning a
notion of witness to each object, and its action on morphisms is the
\emph{fundamental lemma}, that every morphism preserves those
witnesses. Building such a section---by invoking the eliminator---is
thus a semantic interface for constructing a logical relation, with
each component corresponding to a familiar one.

\looseness=-1 A displayed category $\disp{C}$ over $\BiCCC{\Sigma}$ has \emph{displayed
bicartesian closed structure} when it carries displayed terminal and
initial objects, products, coproducts, and exponentials, each over
the corresponding structure in $\BiCCC{\Sigma}$, with displayed
$\beta$/$\eta$ laws. We then call it a \emph{displayed bicartesian closed
category} (\cref{def:dispunivelt,thm:vert-cc}). Just as a model is a functor into such a category, a
displayed model is a section of a displayed category.

\begin{definition}\label{def:dispmodel}
  Given a displayed
  bicartesian closed category $\disp{C}$ over $\BiCCC{\Sigma}$ equipped
  with a displayed interpretation
  $\dispiota : \interpIn{\Sigma}{\disp{C}}$,
  a \emph{displayed model} of $\langsig{\Sigma}$ is
  a structure-preserving section
  $\BiCCC{\Sigma} \to \disp{C}$.
\end{definition}

The eliminator constructs this section from generating data, just as the
recursor constructs a functor from an interpretation.

\begin{theorem}[Elimination principle~\AgdaRef{thm:elim}]\label{thm:elim}
  Let $\disp{C}$ be a displayed bicartesian closed category over
  $\BiCCC{\Sigma}$ with a \emph{displayed interpretation}
  $\dispiota : \interpIn{\Sigma}{\disp{C}}$---a displayed object over $X$
  for each base type $X \in \Sigma_0$ and a displayed morphism over $M$
  for each base term $M \in \Sigma_1$. Then there is a displayed model
  $\eliminterp{\dispiota}$, the section sending each connective of
  $\BiCCC{\Sigma}$ to the chosen displayed structure and each generator
  to its displayed interpretation (\cref{fig:elim}, left).
\end{theorem}

Note, we use products, sums, and functions as illustrative
examples, but nothing in \cref{thm:elim} is specific to this choice of
connectives. Every free category over a signature has a corresponding eliminator
of the same shape: the free cartesian, cartesian closed, and monoidal
categories each take as input a displayed category with the
corresponding displayed structure and a displayed interpretation of
the generators, and each produces a section. The formalization
provides these eliminators uniformly, including for free categories
with an arbitrary selection of universal properties
(\cref{sec:applications}). When we write ``the cartesian
eliminator'' or ``the eliminator for $\Mon{X}$'' throughout, we mean these
instances.

The sense in which $\disp{C}$ carries a displayed version of
bicartesian closed structure is made precise in \cref{sec:dispuniv},
which develops the theory of displayed and vertical universal
properties. \Cref{sec:promotion} shows how the displayed structure
required here is assembled from simpler fiberwise data.

Two special cases of \cref{thm:elim} recover familiar principles:
the recursor and an induction principle. The recursor is the case
where the displayed category does not depend on the syntax. Any
category $\cat{C}$ gives a displayed category over $\BiCCC{\Sigma}$
whose displayed objects and morphisms are those of $\cat{C}$, with no
dependence on the type or term below them. When $\cat{C}$ is
bicartesian closed, this displayed category carries the displayed
structure, and an interpretation of $\Sigma$ in $\cat{C}$ is exactly a
displayed interpretation in it. The displayed model is a section
sending every type to an object of $\cat{C}$ and every term to a
morphism, respecting all the structure---exactly a structure-preserving
functor $\BiCCC{\Sigma} \to \cat{C}$. The eliminator specializes to the
recursor $\recinterp{\iota}$.

\widowpenalty=10000 \looseness=-1 Induction is the case of a \emph{propositional} displayed category,
whose types of displayed objects and morphisms are all propositions.
Such a category is a property of the syntax closed under the rules of
$\lang$, and its displayed model is a section proving that every type
and term of $\langsig{\Sigma}$ satisfies the property---the induction
principle. The general eliminator produces \emph{data} over the
syntax, generalizing both.

\widowpenalty=9999
\subsection{Gluing for Canonicity}
\label{sec:gluing}

We first apply the elimination principle to prove
\emph{canonicity} by categorical gluing. The argument is a
displayed-category presentation of Hermida's fibrational treatment of
logical predicates~\cite{hermida1993fibrations}: Hermida identified
logical predicates with fibrations carrying categorical structure
over the syntax, and obtained the fundamental theorem of logical
relations from that structure. Our approach adapts Hermida's work in
a way amenable to formalization within
intensional type theory in a dependently typed proof assistant:
displayed categories organize the fibrational data as dependently
typed structure over the syntax, with no appeal to equality of
objects (\cref{sec:implementation}). To invoke the eliminator, we must
construct a displayed bicartesian closed category over $\BiCCC{\Sigma}$
whose displayed model encodes the property we are after.

Canonicity says that every closed term is equal to a
\emph{canonical form}: a boolean to $\tmtt$ or $\tmff$, a
number to a numeral. In essence, canonicity states that the equational
theory of the language is sufficient to specify the behavior of an
interpreter.

The logical relation interprets each type $A$ as a family of sets
indexed by the closed terms of $A$: for each closed term
$\cdot \vdash M : A$, a set $\Canonical_{A}\,M$ whose elements are
canonical forms for $M$. The desired section assigns to every closed
term $M : A$ an element of $\Canonical_{A}\,M$, so the section is a
canonicalization function. The relation is \emph{proof-relevant}: it
attaches to each term a set of witnesses rather than a mere truth
value.

The displayed category $\SetD$ (\cref{def:setd}) has such families
as its displayed objects, but it lies over $\Set$ rather
than over the syntax. Reindexing (\cref{def:reindex}) bridges the
gap. The functor
$\Cl\prn{\cdot} \coloneqq \cathomraw{\BiCCC{\Sigma}}{1}{-} :
\BiCCC{\Sigma} \to \Set$ sends each type to its set of closed terms,
and reindexing $\SetD$ along it yields a displayed category over
$\BiCCC{\Sigma}$ whose displayed objects over $A$ are families
indexed by $\Cl\prn{A}$. The reindexed category
$\reindex{\Cl\prn{\cdot}}{\SetD}$ is the setting for the canonicity
logical relation. The displayed category is the type signature of the
logical predicates: a displayed object over $A$ is a predicate on the
closed terms of $A$. A section of it supplies the rest: its object map
is the \emph{relation}, the choice of a predicate over each type, and
its action on terms is the \emph{fundamental lemma}, the proof that
every term respects those predicates. This reindexing is also the
categorical \emph{gluing} construction: gluing along a functor is
reindexing a displayed category along it.

To complete the proof, the reindexed category must be a displayed
bicartesian closed category and must interpret
the generators of $\Sigma$. The
universal structures come with the library, and \cref{sec:dispuniv}
develops them in detail: $\SetD$ satisfies a generalized form of
each of the bicartesian closed universal properties of $\Set$, this
generalized structure is preserved by reindexing, and it yields the
displayed structure that \cref{def:dispmodel} requires.

We must now interpret the generators of
$\Sigma$ as families indexed by closed terms.
If $\Sigma$ has base types for booleans and natural numbers,
we must choose an interpretation for each by defining their canonical forms.
In this case, the canonical forms are the terms in the
image of the following maps from semantic values into syntactic terms:
\begin{align*}
  \QuoteBool\; b &:= \text{if } b \text{ then } \tmtt \text{ else } \tmff,
    & \QuoteNat\; 0 &:= \tmz, \\
  && \QuoteNat\; \prn{\suc\; n} &:= \tms\prn{\QuoteNat\; n}.
\end{align*}
The canonical forms of a closed term $e$ of type $\bool$ or
$\nat$ are then the proofs that $e$ is in the image of the
corresponding quote map: the boolean is either true or false, and
the natural is a numeral. Additionally, we must show that
the interpretations of the constructors $\{\tmtt, \tmff, \tmz, \tms\}$
respect these canonical forms. Having interpreted $\Sigma$ in
$\reindex{\Cl\prn{\cdot}}{\SetD}$, we invoke the eliminator
(\cref{thm:elim}) to obtain the displayed model $\eliminterp{\dispiota}$,
a section over the syntax.

\begin{theorem}[Canonicity~\AgdaRef{thm:canonicity}]\label{thm:canonicity}
  Every closed term $\cdot \vdash M : \bool$ is equal, in the equational
  theory, to $\tmtt$ or $\tmff$, and every closed term
  $\cdot \vdash M : \nat$ is equal to a numeral.
\end{theorem}

\subsubsection{Unwinding the Proof}
\label{sec:gluing:unwinding}
The section given by $\eliminterp{\dispiota}$ describes, for all syntactic
types in $\BiCCC{\Sigma}$, the construction of a canonical form
inductively. It is what we would expect had we constructed the
logical relation by hand. Consider the product type
$\bool \times \nat$: intuitively, a canonical form
of a pair should be a pair of canonical forms for its components.
Indeed, the eliminator strictly preserves products, so for a closed
term $e : \bool \times \nat$, the section returns a
proof that $\pi_1\prn{e}$ is a canonical boolean and
$\pi_2\prn{e}$ is a canonical natural: \(
  \eliminterp{\dispiota}\; e \;:\;
  \Canonical_{\bool}\prn{\pi_1\prn{e}} \times
  \Canonical_{\nat}\prn{\pi_2\prn{e}}
\).
Sum types are similar. For a closed term $e : \bool + \nat$, the
section returns a canonical boolean $e'$ with a proof that
$e = \inl~e'$, or a canonical natural $e''$ with a proof that
$e = \inr~e''$. The disjunction is proof-relevant: an element of it
consists of the tag together with the witness.

The same machinery handles function types, though the construction is
more intricate. Consider a closed term
$f : \bool \Rightarrow \nat$. A canonical form of a
function should send canonical inputs to canonical outputs, so the
section returns a proof that for every closed boolean $e$ with a
canonical form, the application $\app\prn{f, e}$ is a numeral: \(
  \eliminterp{\dispiota}\; f \;:\;
  \PiTy{e}{\Cl\prn{\bool}}{
    \Canonical_{\bool}\prn{e} \to
    \Canonical_{\nat}\prn{\app\prn{f, e}}}
\).
Unlike the product case, this clause is \emph{not} computed
pointwise: it quantifies over all closed arguments $e$ and forms a
function space between the resulting predicates. These two
operations---a quantifier ranging over arguments in the base and a
fiberwise function space---are exactly the \emph{universal
quantifier} and \emph{vertical exponential} developed in
\cref{sec:dispuniv}, combined by reindexing along the application
morphism. The
displayed exponential that the eliminator uses for
$\bool \Rightarrow \nat$ is assembled from these pieces.

The section acts on open terms as well. A term
$x : A \vdash M : B$ is a morphism of $\BiCCC{\Sigma}$, and the
section sends it to a map of families: given a closing
substitution---a closed term $N : A$---and a canonical form for $N$,
it produces a canonical form for $M[N/x]$. Canonicity for closed
terms is the special case of the unit context.

Note, the user only specifies the logical relation at base types, and
the value of the relation at each type connective is determined by its
universal property. Because the relation is proof-relevant, the
section computes canonical forms instead of only proving that they
exist. Evaluated at a closed boolean $M$, the section yields a value
$b$ together with a proof that $\QuoteBool\;b = M$ in the equational
theory: an interpreter whose correctness proof is part of its output.

The same principle holds for additional connectives. Each new type
former corresponds to a new universal property, and extending the
language requires only that the model satisfy that property.

\section{Logical Relations via Displayed Category Theory}
\label{sec:dispuniv}

The elimination principle of \cref{sec:free-cats:elim} constructs a
displayed model---a section over the syntax---from a displayed
bicartesian closed category with an interpretation, and this section
encodes a metatheoretic result. What remains is to construct those
displayed categories. This section develops the necessary machinery: a
uniform presentation of universal properties as universal elements,
their displayed and vertical generalizations, and the reindexing and
promotion theorems that assemble them from reusable parts. The
displayed category of sets $\SetD$, the building block behind the
canonicity proof of \cref{sec:gluing}, is threaded through the section
as the demonstrative case: we instantiate each concept in $\SetD$ as
it appears. The section ends with \cref{constr:assemble}, which
modularly assembles displayed models connective-by-connective.
Readers who know fibered category theory may find
these ideas familiar, and we compare the two in \cref{sec:fibered}.

\subsection{Displayed Category Theory Preliminaries}
\label{sec:dispcat-prelim}

Displayed category theory is the category-theoretic analogue of
dependent type theory. A category plays the role of a context, and a
displayed category $\disp{C}$ over $\cat{C}$ that of a dependent type
in the context $\cat{C}$: to each object $c$ it assigns a type of
displayed objects depending on $c$, and to each morphism a type of
displayed morphisms, while a section of $\disp{C}$ is a dependent
function, selecting a displayed object over every object and a
displayed morphism over every morphism. This analogy drives our
use of displayed categories, and it extends to every construction
below. We recall the total
category, reindexing, and the displayed category of sets $\SetD$.

\begin{definition}[{\cite[Definition 3.2]{ahrens2019displayed}}~\AgdaRef{def:totalcat}]
  \label{def:totalcat}
  Let $\disp{C}$ be a displayed category over a category $\cat{C}$. The
  \emph{total category} $\totalcat{\disp{C}}$ is the category whose
  objects are pairs $\prn{c, \mkdisp{c}}$ with $\inob{c}{C}$ and
  $\indispob{\mkdisp{c}}{C}{c}$, and whose morphisms from
  $\prn{c, \mkdisp{c}}$ to $\prn{d, \mkdisp{d}}$ are pairs
  $\prn{f, \mkdisp{f}}$ with $\incathom{f}{C}{c}{d}$ and
  $\mkdisp{f} \in \disphom{C}{f}{\mkdisp{c}}{\mkdisp{d}}$.
\end{definition}

Just as context extension produces a new context $\Gamma, x : A$ from
$\Gamma$ and a type $A$, the total category extends the base $\cat{C}$
by the displayed category $\disp{C}$---it is a \emph{context former}. It
comes equipped with a projection functor
$\pi : \totalcat{\disp{C}} \to \cat{C}$. Displayed structure on $\disp{C}$ amounts to the total category
$\totalcat{\disp{C}}$ possessing those universal properties, preserved
strictly by $\pi$.

\begin{definition}[{\cite[Definition 3.15]{ahrens2019displayed}}~\AgdaRef{def:reindex}]
  \label{def:reindex}
  Given a functor $F : \cat{B} \to \cat{C}$ and a displayed category
  $\disp{C}$ over $\cat{C}$, the \emph{reindexing} of $\disp{C}$
  along $F$, written $\reindex{F}{\disp{C}}$, is the displayed
  category over $\cat{B}$ whose fiber over $b$ is the fiber of
  $\disp{C}$ over $F\prn{b}$. That is,
  $\fiberofexpr{\reindex{F}{\disp{C}}}{b} := \dispob{C}{F\prn{b}}$.
  A displayed morphism of $\reindex{F}{\disp{C}}$ over
  $g : \cathom{B}{b}{b'}$ is a
  displayed morphism of $\disp{C}$ over $F\prn{g}$, that is
  \looseness=-1 $\disphom{\reindex{F}{\disp{C}}}{g}{\mkdisp{b}}{\mkdisp{b'}} :=
  \disphom{C}{F\prn{g}}{\mkdisp{b}}{\mkdisp{b'}}$.
\end{definition}

Reindexing is the analogue of substitution: substituting along a function
$f : A' \to A$ turns a dependent type $B$ over $A$ into the dependent
type $B[f]$ over $A'$, and reindexing along a functor
$F : \cat{B} \to \cat{C}$ does the same to a displayed category
over $\cat{C}$. Reindexing lets us work
compositionally. Rather than building a displayed category over the
syntax directly, we construct displayed categories over familiar
semantic categories once, then reindex them onto the syntax at hand.
For canonicity, we reindex a category of displayed sets along the
closed-terms functor (\cref{sec:gluing}). For Kripke logical
relations, we reindex displayed presheaves along a nerve
(\cref{sec:conservativity}).

\looseness=-1 The simplest semantic domain is the displayed category of sets
$\SetD$ (\cref{def:setd}). Its structure is computed pointwise, suiting
it to canonicity, where the logical relation is a family over closed
terms.

\subsection{Universal Properties}
\label{sec:univelts}

A universal property is the semantic essence of a type former. It
specifies the maps into (or out of) an object, together with their
$\beta$/$\eta$ laws, and determines the object up to
unique isomorphism. A presheaf records the
data that a prospective object must obey, and an object satisfies
the universal property exactly when it \emph{represents} the
presheaf. We recall the notions of representability and universal
element, following Riehl~\cite[\S2.2--2.3]{riehl2017category}.

A \emph{presheaf} on a category $\cat{C}$ is a functor
$P : \op{C} \to \Set$. The \emph{Yoneda embedding} sends an object
$c$ to the presheaf $\yo{c}$, defined by
$\yo{c}\prn{d} = \cathom{C}{d}{c}$.

\begin{definition}
  \label{def:elt}
  An \emph{element} of a presheaf $P$ is a pair $\elt{c}{p}$ of an object $\inob{c}{C}$ and a point
  $p \in \pshelt{P}{c}$.
\end{definition}

By the Yoneda lemma, each element $\elt{c}{p}$ of $P$ determines a
natural transformation $\recyo{c, p} : \yo{c} \to P$, defined pointwise
by $\recyo{c, p}\prn{f : d \to c} = \pshelt{P}{f}\prn{p}$. Given a morphism
$\incathom{f}{C}{d}{c}$, this takes the action of $P$ on $f$, a
function $\pshelt{P}{c} \to \pshelt{P}{d}$, and passes the point $p$
through it. We read $\recyo{c, p}$ as composition of $f$ with the
point $p$. Indeed, when $P$ is itself of the form $\yo{c'}$,
$\recyo{c, p}$ is literally the precomposition action on morphisms.

\begin{definition}
  \label{def:univelt}
  An element $\elt{c}{p}$ is \emph{universal} if $\recyo{c, p}$ is a
  natural isomorphism. A presheaf is \emph{representable} if it
  admits a universal element---equivalently, if it is isomorphic to
  \looseness=-1 some $\yo{c}$.
\end{definition}

We work with universal elements rather than representability throughout
the formalization. Representability requires exhibiting a natural
isomorphism $P \cong \yo{c}$: a family of bijections together with a
proof of naturality. A universal element is more compact---an
object $c$ and a point $p \in \pshelt{P}{c}$---and it replaces the
naturality obligation with a pointwise bijectivity condition on
$\recyo{c, p}$. The Yoneda lemma guarantees naturality. This reduction in proof
obligations makes universal elements the more convenient interface for
constructing and composing universal properties in our development.

The notion of universal
element provides a single interface for reasoning about universal
properties in general, which is advantageous for code reuse throughout
our mechanization. Further, this generic treatment of universal
properties scales to the setting of displayed categories
(\cref{sec:dispunivprops}). Because we uniformly describe universal
constructions, we may reuse generic theorems when reasoning about
them.

\paragraph{Universal properties as universal elements} Each universal
property in our development
is formulated as a universal element of an appropriate
presheaf. As an example, consider binary products. Given objects $a, b$
in $\cat{C}$, the \emph{product presheaf} of $\yo{a}$ and $\yo{b}$
sends each object $c$ to the
set $\cathom{C}{c}{a} \times \cathom{C}{c}{b}$ of pairs of morphisms. A
universal element of this presheaf is an object $a \times b$ together with
a pair of morphisms $\pi_1 : a \times b \to a$ and
$\pi_2 : a \times b \to b$ such that for every $c$, the map
$f \mapsto \prn{\seq{f}{\pi_1},\, \seq{f}{\pi_2}}$ is a bijection. This
is exactly the universal property of the product $a \times b$.

\subsection{Displayed Universal Properties}
\label{sec:dispunivprops}

We now generalize universal properties to the displayed setting.
\Cref{sec:univelts} reduced universal properties to universal
elements of presheaves, so the generalization requires displayed
counterparts of the same apparatus: \emph{displayed presheaves},
their elements, and a displayed Yoneda theory characterizing when an
element is universal. With the apparatus in place, displayed
\looseness=-1 universal properties come in two variants. A \emph{displayed}
universal property lies over a chosen universal property in the base;
it is the form of structure a displayed model requires
(\cref{def:dispmodel}). A \emph{vertical} universal property has
its defining data in a single fiber; it is the form that reindexing
preserves, and therefore the form in which we verify structure on
reusable semantic components. The universal property of a vertical
element constrains displayed morphisms from all fibers, not only its
own, and \cref{sec:promotion} converts vertical structure into
displayed structure. Both are notions of \emph{indexed universal
properties}. While an ordinary universal property specifies a unique
object in a category, these specify unique displayed objects within a
displayed category. The universal displayed object is tied to the
corresponding universal object in the base, and the universal maps
into or out of it are displayed morphisms lying over the universal
maps into or out of that base object.

\subsubsection{Displayed Presheaves}
\label{sec:disppsh}

A displayed presheaf generalizes a presheaf to account for displayed
structure. Let $\disp{C}$ be a displayed category over $\cat{C}$ and
$P$ a presheaf on $\cat{C}$: where a presheaf on $\cat{C}$ assigns a
set to each object, a displayed presheaf $\disppsh{P}$ over $P$ must
additionally depend on displayed objects and elements of $P$. Concretely, a displayed
presheaf $\disppsh{P}$ assigns to each object $\inob{c}{C}$, displayed
object $\indispob{\mkdisp{c}}{C}{c}$, and element $p \in P\,c$ a set
$\disppshelt{P}{\mkdisp{c}}{p}$, functorially in the triple
$(c, \mkdisp{c}, p)$. The following definitions package the triples
into a single category, so that functoriality can be stated once.

\begin{definition}[\AgdaRef{def:catofelts}]
  \label{def:catofelts}
  Let $P$ be a presheaf on $\cat{C}$. The \emph{category of
  elements}~\cite[\S2.4]{riehl2017category} of
  $P$, written $\catofelts{P}$, is the displayed category over $\cat{C}$
  with displayed objects over $c$ the elements $p \in P\,c$, and
  displayed morphisms over $\incathom{f}{C}{c}{d}$ from $p$ to $q$ the
  proofs that $(Pf)\,q = p$.
\end{definition}

\begin{definition}[\AgdaRef{def:slicecat}]
  \label{def:slicecat}
  Let $\disp{C}$ be a displayed category over $\cat{C}$ and $P$ a
  presheaf on $\cat{C}$. The \emph{slice category}
  $\slicecat{\disp{C}}{P}$ is the displayed category over $\cat{C}$
  obtained as the fiberwise product of $\disp{C}$ and $\catofelts{P}$.
  Its objects over $c$ are pairs $(\mkdisp{c}, p)$ with
  $\indispob{\mkdisp{c}}{C}{c}$ and $p \in P\,c$.
\end{definition}

\begin{definition}[\AgdaRef{def:displayedpsh}]
  \label{def:displayedpsh}
  A \emph{displayed presheaf} $\disppsh{P}$ over $P$ is a presheaf on
  the total category $\totalcat{(\slicecat{\disp{C}}{P})}$.
\end{definition}

The prototypical example of a displayed presheaf is the \emph{displayed
representable}.

\begin{definition}[\AgdaRef{def:dispyo}]
  \label{def:dispyo}
  Given a displayed object $\indispob{\mkdisp{c}}{C}{c}$, the
  \emph{displayed representable} $\dispyo{\mkdisp{c}}$ is the displayed
  presheaf over the representable $\yo{c}$ sending $(\mkdisp{x}, f)$ to
  the set of displayed morphisms
  $\disphom{C}{f}{\mkdisp{x}}{\mkdisp{c}}$.
\end{definition}

\looseness=-1 The assignment $\mkdisp{c} \mapsto \dispyo{\mkdisp{c}}$ is the displayed
analogue of the Yoneda embedding: it sends displayed objects to
displayed presheaves over representables. We expect a displayed
counterpart to this principle: a displayed
element of $\disppsh{P}$ ought to induce a map out of the displayed
representable into $\disppsh{P}$, just as an ordinary element induces a map out of $\yo{c}$.
Stating it is not immediate, because the two displayed presheaves
sit over different bases---the representable $\yo{c}$ and
$P$---linked by a morphism $\yo{c} \to P$. Expressing a map of
displayed presheaves over such a base morphism requires pulling the
target back along it: a morphism $\alpha : P \to Q$ pulls a displayed
presheaf over $Q$ back to one over $P$.

\begin{definition}[\AgdaRef{def:reindexdisppsh}]
  \label{def:reindexdisppsh}
  Let $\alpha : P \to Q$ be a morphism of presheaves on $\cat{C}$ and
  $\disppsh{Q}$ a displayed presheaf over $Q$. The morphism $\alpha$
  induces a functor
  $\totalcat{(\slicecat{\disp{C}}{P})} \to
  \totalcat{(\slicecat{\disp{C}}{Q})}$, sending a triple
  $(c, \mkdisp{c}, p)$ to $(c, \mkdisp{c}, \alpha_c\,p)$. The
  \emph{reindexed presheaf} $\cartlift{\alpha}{\disppsh{Q}}$ over $P$ is
  $\disppsh{Q}$ precomposed with this functor.
\end{definition}

A displayed presheaf over $P$ is an ordinary presheaf on
$\totalcat{(\slicecat{\disp{C}}{P})}$, so a morphism between two of them
is simply a morphism of those presheaves.

\begin{definition}[\AgdaRef{def:disppshmor}]
  \label{def:disppshmor}
  A \emph{morphism of displayed presheaves} $\disppsh{P} \to
  \disppsh{Q}$ over the same base $P$ is a morphism of presheaves on
  $\totalcat{(\slicecat{\disp{C}}{P})}$. More generally, given a map
  of presheaves $\alpha : P \to Q$, a morphism $\disppsh{P} \to
  \disppsh{Q}$ over $\alpha$ is a morphism from $P$ into the reindexed
  presheaf $\cartlift{\alpha}{\disppsh{Q}}$.
\end{definition}

With maps of displayed presheaves over a base morphism now available, we
can state the displayed Yoneda lemma, as in the ordinary case, to
\looseness=-1 say when a displayed element is universal.

\subsubsection{Displayed Universal Elements}
\label{sec:dispunivprops:elts}

The central notion our framework is built on is the \emph{displayed
universal element}. A displayed universal property is a universal
element valued in a displayed category, lying strictly over a
universal property in the base. It specifies how a logical relation
interprets a type constructor.
The eliminator of
\cref{sec:free-cats:elim} requires displayed terminal and initial
objects, products, coproducts, and exponentials lying over their base
counterparts, and each such structure is a displayed universal
element of an appropriate displayed presheaf.

\looseness=-1 The morphisms of \cref{def:disppshmor} let us characterize when a
displayed element is universal.

\begin{definition}
  \label{def:dispelt}
  A \emph{displayed element} of $\disppsh{P}$ over an element
  $\elt{c}{p}$ of $P$ is a pair $\dispelt{\mkdisp{c}}{\mkdisp{p}}$
  consisting of a displayed object $\indispob{\mkdisp{c}}{C}{c}$ and a
  point $\mkdisp{p} \in \disppshelt{P}{\mkdisp{c}}{p}$.
\end{definition}

Just as the Yoneda lemma sends each element of $P$ to a map out of a
representable, each displayed element
$\dispelt{\mkdisp{c}}{\mkdisp{p}}$ of $\disppsh{P}$ over
$\elt{c}{p}$ determines a morphism of displayed presheaves
$\recdispyo{\mkdisp{c}, \mkdisp{p}} : \dispyo{\mkdisp{c}} \to
\disppsh{P}$ lying over $\recyo{c, p} : \yo{c} \to P$, defined
pointwise by $\recdispyo{\mkdisp{c}, \mkdisp{p}}\prn{\mkdisp{f}} =
\disppshelt{P}{\mkdisp{f}}{\mkdisp{p}}$; this is the map
anticipated in \cref{sec:disppsh}. The assignment is a bijection
between displayed elements over $\elt{c}{p}$ and morphisms of
displayed presheaves $\dispyo{\mkdisp{c}} \to \disppsh{P}$ over
$\recyo{c, p}$---the \emph{displayed Yoneda lemma}.

\begin{definition}[\AgdaRef{def:dispunivelt}]
  \label{def:dispunivelt}
  A displayed element $\dispelt{\mkdisp{c}}{\mkdisp{p}}$ over a
  universal element $\elt{c}{p}$ of $P$ is a \emph{displayed universal
  element} if
  $\recdispyo{\mkdisp{c}, \mkdisp{p}}$ is an isomorphism of displayed
  presheaves.
\end{definition}

\clubpenalty=10000 Equivalently, a displayed universal element assembles into an ordinary
universal element of the total category $\totalcat{\disp{C}}$ lying
\emph{strictly} over the corresponding universal element in $\cat{C}$:
first projection sends it to $\elt{c}{p}$ on the nose, not up to a path.
This strictness is why displayed categories are a distinct notion---an
equality of objects imposed after the fact cannot be strictified---so
the theory places the displayed data over the base by construction, and
the eliminator preserves type structure exactly rather than up to
isomorphism.

\clubpenalty=9999
\emph{Displayed} products, exponentials, coproducts, and so on are
displayed universal elements over the chosen universal elements of the
base---equivalently, structure on the total category that $\pi$
preserves strictly. A displayed category with all the displayed
structure of a cartesian closed category is \emph{displayed cartesian
closed}, likewise for bicartesian closed and monoidal structure. For products, the
displayed presheaf in question lies over the product presheaf of
\cref{sec:univelts}: it is the product of the displayed
\looseness=-1 representables $\dispyo{\mkdisp{a}}$ and $\dispyo{\mkdisp{b}}$, each
reindexed along its projection (\cref{def:reindexdisppsh}).

A displayed category with these displayed universal properties carries
exactly the structure the elimination principle of
\cref{sec:free-cats:elim} takes as input. Displayed universal
properties are not, however, the form in which that structure is best
established. A displayed universal property lies over one in the
base, and reindexing need not preserve base universal properties. The
vertical universal element, developed next, makes no reference
to universal properties of the base. It survives reindexing along any
functor, and \cref{sec:promotion} recovers displayed structure from
it.

\subsubsection{Vertical Universal Elements}

A vertical presheaf is a displayed presheaf whose base is a
representable $\yo{c}$, and this representable presheaf has a simple
universal element; namely, the identity $\id_c$. Therefore,
\emph{vertical} universal elements are only specifying structure on the
displayed level, because we already know which element in the base
category they are defined over.

\begin{definition}[\AgdaRef{def:verticalpsh}]
  \label{def:verticalpsh}
  A \emph{vertical presheaf} $\vertpsh{P}$ at $\inob{c}{C}$ is a
  displayed presheaf over $\yo{c}$.
\end{definition}

Concretely, since $\yo{c}\prn{x} = \cathom{C}{x}{c}$, the action of
$\vertpsh{P}$ takes a displayed object $\indispob{\mkdisp{x}}{C}{x}$
together with a morphism $\incathom{f}{C}{x}{c}$ into the object $c$ and
returns a set $\vertpsh{P}\prn{\mkdisp{x}, f}$. The displayed
representable lands among vertical presheaves: $\dispyo{\mkdisp{c}}$ is a
vertical presheaf at $c$, with
$\dispyo{\mkdisp{c}}\prn{\mkdisp{x}, f} = \disphom{C}{f}{\mkdisp{x}}{\mkdisp{c}}$.

Consider the displayed elements (\cref{def:dispelt}) of a vertical
presheaf. An element of the base $\yo{c}$ is a morphism
$\incathom{g}{C}{x}{c}$, so a displayed element of $\vertpsh{P}$ over $g$
is a displayed object $\indispob{\mkdisp{x}}{C}{x}$ together with a point
in the fiber $\vertpsh{P}\prn{\mkdisp{x}, g}$. Among these, the
elements lying over the \emph{identity}, the
universal element of $\yo{c}$, are distinguished. A displayed element over
$\id_c$ has its vertex over $c$ itself.

\begin{definition}[\AgdaRef{def:vertunivelt}]
  \label{def:vertunivelt}
  A \emph{vertical element} of a vertical presheaf $\vertpsh{P}$ at $c$
  is a displayed object $\indispob{\mkdisp{c}}{C}{c}$ over $c$
  together with a point
  $\mkdisp{e} \in \vertpsh{P}\prn{\mkdisp{c}, \id_c}$ in the fiber
  over the identity. It determines a map of
  vertical presheaves
  $\recvertyo{\mkdisp{e}} : \dispyo{\mkdisp{c}} \to \vertpsh{P}$,
  sending a
  displayed morphism $\mkdisp{f}$ into the vertex to an element of
  $\vertpsh{P}\prn{\mkdisp{x}, f}$. The
  \looseness=-1 vertical element is a \emph{vertical universal element} if
  $\recvertyo{\mkdisp{e}}$ is an isomorphism.
\end{definition}

The universal property in \cref{def:vertunivelt} reaches across fibers.
Even though the vertical element $(\mkdisp{c}, \mkdisp{e})$ lives over
the identity, the isomorphism $\recvertyo{\mkdisp{e}}$ asserts that for
\emph{every} displayed object $\indispob{\mkdisp{x}}{C}{x}$ and
\emph{every} morphism $\incathom{f}{C}{x}{c}$, the displayed morphisms
$\disphom{C}{f}{\mkdisp{x}}{\mkdisp{c}}$ into the vertex are in
bijection with the elements of $\vertpsh{P}\prn{\mkdisp{x}, f}$. Because
this ranges over displayed morphisms over arbitrary base
morphisms into $c$, and not only the \emph{vertical} morphisms---the
displayed morphisms over $\id_c$---a vertical universal property is
stronger than the corresponding structure
in the fiber over $c$ alone.

\begin{intensionalremark}
  Extensionally, a vertical universal element is a displayed
  universal element over $\id_c$. In the formalization the
  two are distinct notions with a conversion between them.
  \Cref{sec:implementation:strictness} explains the reason, which is
  definitional versus propositional equality of fibers, and the
  strict-base situation in which the distinction collapses.
\end{intensionalremark}

\subsubsection{Examples of Vertical Universal Elements}

Several examples of vertical universal elements
recur throughout the development, each with a specific role in
the canonicity relation of \cref{sec:gluing}: vertical terminals and
products interpret the unit and product types once promoted
(\cref{sec:promotion}), cartesian lifts implement substitution, and
vertical exponentials appear in the function-type clause. Throughout, each displayed category
$\disp{C}$ over $\cat{C}$ determines a \emph{fiber category}
$\fibercat{C}{c}$ over each object $c$, whose objects are the displayed
objects $\dispob{C}{c}$ and whose morphisms are the vertical morphisms.

The simplest example of a vertical universal element is given by vertical
terminal objects.
\begin{definition}[\AgdaRef{def:vertterm}]
  \label{def:vertterm}
  A displayed category $\disp{C}$ over $\cat{C}$ has \emph{vertical
    terminal objects} if for each $\inob{c}{C}$, there is a vertical
  universal element of the terminal presheaf at $c$.
  The vertical terminal over $c$ admits, for every displayed
  object $\indispob{\mkdisp{d}}{C}{d}$ and every morphism
  \looseness=-1 $\incathom{f}{C}{d}{c}$, a unique displayed morphism from $\mkdisp{d}$
  to it over $f$.
\end{definition}

Binary products are similar. Given
$\mkdisp{x}, \mkdisp{y} : \dispob{C}{c}$, the vertical product
presheaf sends a displayed object $\indispob{\mkdisp{d}}{C}{d}$ and a
morphism $\incathom{f}{C}{d}{c}$ to the set of pairs of displayed
morphisms $\disphom{C}{f}{\mkdisp{d}}{\mkdisp{x}} \times
\disphom{C}{f}{\mkdisp{d}}{\mkdisp{y}}$. That is, the presheaf is
the product $\dispyo{\mkdisp{x}} \times \dispyo{\mkdisp{y}}$ of the
displayed representables.

\begin{definition}[\AgdaRef{def:vertprod}]
  \label{def:vertprod}
  A \emph{vertical product} of $\mkdisp{x}, \mkdisp{y} : \dispob{C}{c}$
  is a vertical universal element of
  $\dispyo{\mkdisp{x}} \times \dispyo{\mkdisp{y}}$: a displayed object
  $\vertprod{\mkdisp{x}}{\mkdisp{y}}$ over $c$
  such that every displayed morphism into
  $\vertprod{\mkdisp{x}}{\mkdisp{y}}$, over any $\incathom{f}{C}{d}{c}$,
  factors uniquely as a pair of displayed morphisms into $\mkdisp{x}$
  and $\mkdisp{y}$ over $f$.
\end{definition}

\begin{definition}[\AgdaRef{def:vertbp}]
  \label{def:vertbp}
  A displayed category $\disp{C}$ over $\cat{C}$ has \emph{vertical
    binary products} if a vertical product (\cref{def:vertprod}) exists
  for each $\inob{c}{C}$ and each pair
  $\mkdisp{x}, \mkdisp{y} : \dispob{C}{c}$ in a common fiber.
\end{definition}

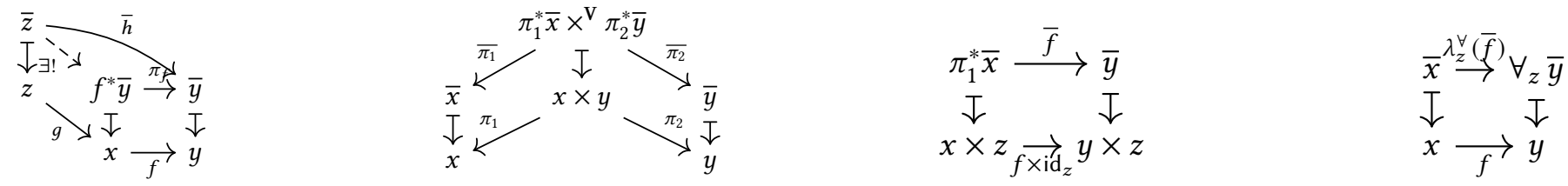
\begin{figure}
\centering
\begin{minipage}[b]{0.25\linewidth}
\centering
\scalebox{0.6}{%
\begin{tikzcd}[ampersand replacement=\&, sep=small]
  \mkdisp{z} \ar[dr, dashed, "\exists!"'] \ar[drr, bend left=15, "\mkdisp{h}"] \ar[d, mapsto] \& \& \\
  z \ar[dr, "g"'] \& \cartlift{f}{\mkdisp{y}} \ar[r, "\pi_f"] \ar[d, mapsto]
    \& \mkdisp{y} \ar[d, mapsto] \\
  \& x \ar[r, "f"'] \& y
\end{tikzcd}}
\par\smallskip{\footnotesize (a)}
\end{minipage}\hfill
\begin{minipage}[b]{0.25\linewidth}
\centering
\scalebox{0.6}{%
\begin{tikzcd}[ampersand replacement=\&, sep=small]
  \& \vertprod{\cartlift{\pi_1}{\mkdisp{x}}}{\cartlift{\pi_2}{\mkdisp{y}}}
    \ar[dl, "\mkdisp{\pi_1}"'] \ar[dr, "\mkdisp{\pi_2}"] \ar[d, mapsto] \& \\
  \mkdisp{x} \ar[d, mapsto] \&
    x \times y \ar[dl, "\pi_1"'] \ar[dr, "\pi_2"] \&
    \mkdisp{y} \ar[d, mapsto] \\
  x \& \& y
\end{tikzcd}}
\par\smallskip{\footnotesize (b)}
\end{minipage}\hfill
\begin{minipage}[b]{0.24\linewidth}
\centering
\scalebox{0.72}{%
\begin{tikzcd}[ampersand replacement=\&, sep=small]
  \cartlift{\pi_1}{\mkdisp{x}} \ar[r, "\mkdisp{f}"] \ar[d, mapsto] \&
    \mkdisp{y} \ar[d, mapsto] \\
  x \times z \ar[r, "f \times \id_z"'] \& y \times z
\end{tikzcd}}
\par\smallskip{\footnotesize (c)}
\end{minipage}\hfill
\begin{minipage}[b]{0.24\linewidth}
\centering
\scalebox{0.72}{%
\begin{tikzcd}[ampersand replacement=\&, sep=small]
  \mkdisp{x} \ar[r, "\lamforall{z}(\mkdisp{f})"] \ar[d, mapsto] \&
    \uq{z}{\mkdisp{y}} \ar[d, mapsto] \\
  x \ar[r, "f"'] \& y
\end{tikzcd}}
\par\smallskip{\footnotesize (d)}
\end{minipage}
\caption{\emph{(a)} the universal property of a cartesian lift;
\emph{(b)} the displayed product as a vertical product of cartesian
lifts. The universal quantifier $\uq{z}{(-)}$, right adjoint to
cartesian lifting along $\pi_1$, expresses the equivalence of
\emph{(c)} displayed morphisms into $\mkdisp{y}$ over $f \times \id_z$
and \emph{(d)} displayed morphisms into $\uq{z}{\mkdisp{y}}$ over $f$.}
\label{fig:cartlift-prod}\label{fig:univquant}
\end{figure}

A second kind of example connects the fibers themselves. A
fibration relates the fiber categories functorially: each morphism
$\incathom{f}{C}{x}{y}$ in the base induces a
functor $f^* : \fibercat{C}{y} \to \fibercat{C}{x}$ that carries
displayed objects and morphisms from the fiber over $y$ to the fiber over
$x$. This functorial structure manifests through \emph{cartesian
  lifts}. A cartesian lift
provides, for each displayed object
$\indispob{\mkdisp{y}}{C}{y}$ and morphism $\incathom{f}{C}{x}{y}$, a
canonical way to interpret $\mkdisp{y}$ in the fiber over $x$. The
cartesian lift $\cartlift{f}{\mkdisp{y}}$ is the interpretation of
$\mkdisp{y}$ pulled back along $f$, with a displayed morphism
$\pi_f : \disphom{C}{f}{\cartlift{f}{\mkdisp{y}}}{\mkdisp{y}}$
\looseness=-1 satisfying the universal property in \cref{fig:cartlift-prod}(a).

Formally, cartesian lifts are vertical universal elements of a reindexed
displayed representable. The displayed representable $\dispyo{\mkdisp{y}}$
is a vertical presheaf \emph{at $y$}. Given $\incathom{f}{C}{x}{y}$,
we pull it back to a vertical presheaf at $x$ along
$\yo{f} : \yo{x} \to \yo{y}$, the Yoneda image of $f$ (postcomposition
with $f$). The reindexed presheaf $\cartlift{(\yo{f})}{\dispyo{\mkdisp{y}}}$
of \cref{def:reindexdisppsh} sends a displayed object
$\indispob{\mkdisp{d}}{C}{d}$ and a morphism $\incathom{g}{C}{d}{x}$ to
the set of displayed morphisms
$\disphom{C}{\seq{g}{f}}{\mkdisp{d}}{\mkdisp{y}}$, the displayed hom
\looseness=-1 into $\mkdisp{y}$ reindexed along $f$.

\begin{definition}
  \label{def:cartlift}\label{def:fibration}
  A \emph{cartesian lift} of $\mkdisp{y}$ along $f$ is a vertical
  universal element of the reindexed displayed representable
  $\cartlift{(\yo{f})}{\dispyo{\mkdisp{y}}}$. Write
  $\cartlift{f}{\mkdisp{y}}$ for the representing displayed object,
  which lives in $\dispob{C}{x}$. $\disp{C}$
  is a \emph{fibration} if every $\indispob{\mkdisp{y}}{C}{y}$ has a
  cartesian lift along every $\incathom{f}{C}{x}{y}$
  (\cite[Definition 5.2]{ahrens2019displayed}).
\end{definition}

In the categorical semantics of type theory, morphisms of the base
are substitutions, and the functor
$f^* : \fibercat{C}{y} \to \fibercat{C}{x}$ models their action on
the types in context.

\begin{theorem}[\AgdaRef{thm:setd-fibration}]\label{thm:setd-fibration}
  \looseness=-1 $\SetD$ is a fibration: the cartesian lift $\cartlift{f}{P}$ of $P$
  along $f : Y \to X$ is $\lambda y.\; P\prn{f\prn{y}}$.
\end{theorem}

A final example combines the two. The vertical exponential plays the
role of the exponential within a fiber; in $\SetD$ it is the pointwise
function space from the function-type clause
\looseness=-1 (\cref{sec:gluing:unwinding}).
\begin{definition}[\AgdaRef{def:vertexp}]
  \label{def:vertexp}
  A displayed category $\disp{C}$ that is a fibration with vertical
  binary products has \emph{vertical exponentials} if for each
  $\inob{c}{C}$ and displayed objects
  $\mkdisp{x}, \mkdisp{y} : \dispob{C}{c}$, there is a vertical
  universal element of the exponential presheaf sending a displayed
  object $\indispob{\mkdisp{d}}{C}{d}$ and a morphism
  $\incathom{f}{C}{d}{c}$ to the set of displayed morphisms
  $\disphom{C}{f}{\vertprod{\mkdisp{d}}{\cartlift{f}{\mkdisp{x}}}}{\mkdisp{y}}$,
  \looseness=-1 with $\cartlift{f}{\mkdisp{x}}$ the cartesian lift of $\mkdisp{x}$
  over $d$. We write $\vertexp{\mkdisp{x}}{\mkdisp{y}}$ for it.
\end{definition}

\subsubsection{Universal Quantifiers}
\label{sec:uq}

The function-type clause of the canonicity relation quantified over
all closed arguments (\cref{sec:gluing:unwinding}). The universal
quantifier is that operation in general, and it is the one piece of
vertical structure that reaches across fibers.

Its name comes from reading the fibration $\disp{C} \to \cat{C}$ as a
hyperdoctrine in the sense of Lawvere~\cite{lawvere1969adjointness}
(see also~\cite{JacobsCLTT}). A displayed object over a base object
$y$ is a proof-relevant predicate in the context $y$, and a morphism
in the base is a substitution between contexts. Under this reading,
cartesian lifting along the projection
$\incathom{\pi_1}{C}{y \times z}{y}$ is \emph{weakening}: it
reinterprets a predicate in context $y$ as one in the extended context
$y \times z$ that ignores the new variable $z$. Lawvere's observation
is that the two quantifiers, when they exist, are precisely the
adjoints to weakening: the existential on the left, the universal on
the right.  The right adjoint $\uq{z}{(-)}$ therefore sends a
predicate in context $y \times z$ to the predicate ``for all $z$,
$\ldots$'' in context $y$, exactly the universal quantifier of
predicate logic.

Let $\cat{C}$ have binary products and let $\disp{C}$ be a fibration
over $\cat{C}$. For any $\inob{z}{C}$, cartesian lifting along $\pi_1$
pulls displayed objects from a fiber over $y$ into the fiber over
$y \times z$. The universal quantifier $\uq{z}{(-)}$ is specified as
the right adjoint to cartesian lifting.

\begin{definition}[\AgdaRef{def:uq}]
  \label{def:uq}
  $\disp{C}$ \emph{has universal quantifiers} if cartesian lifting
  along $\pi_1$ has a right adjoint, $\uq{z}{(-)}$.  For
  $\mkdisp{x} : \dispob{C}{x}$, $\mkdisp{y} : \dispob{C}{y \times z}$,
  and $\incathom{f}{C}{x}{y}$, the displayed morphisms of
  \cref{fig:univquant}(c,\,d) are in natural bijection.
\end{definition}

The introduction form
$\lamforall{z} : \disphom{C}{f \times \id_z}{\cartlift{\pi_1}{\mkdisp{x}}}{\mkdisp{y}}
\to \disphom{C}{f}{\mkdisp{x}}{\uq{z}{\mkdisp{y}}}$
abstracts a displayed morphism into $\mkdisp{y}$ to one into
$\uq{z}{\mkdisp{y}}$. The elimination form
$\indisphom{\appforall{z}}{C}{\id}{\cartlift{\pi_1}{\uq{z}{\mkdisp{y}}}}{\mkdisp{y}}$
\looseness=-1 eliminates the quantifier.

\subsubsection{Duality: Coproducts and Existential Quantifiers}
\label{sec:duality}

Every notion in this section dualizes. A displayed category
$\disp{C}$ over $\cat{C}$ has an opposite $\dispop{C}$ over $\op{C}$,
and the dictionary carries over: vertical \emph{initial objects} in
$\disp{C}$ are vertical terminal objects in $\dispop{C}$, vertical
\emph{coproducts} are vertical products, and \emph{opcartesian
lifts} (the pushforwards $f_!$) are cartesian lifts. The
\emph{existential quantifier} $\exists_z$ of Lawvere's picture arises
alongside: it is the left adjoint to the same weakening functor whose
right adjoint is the universal quantifier (\cref{sec:uq}). Sum and empty types need only the first three of these:
vertical initial objects, vertical coproducts, and opcartesian lifts,
which together with vertical cartesian closed structure
(\cref{def:vertcc}, defined next) bundle into \emph{vertically
bicartesian closed} structure. The existential quantifier is not
required, because the language has no dual of function types. In
$\SetD$ the dual structure is again pointwise: vertical coproducts
are pointwise disjoint unions, and opcartesian lifts are dependent
pair types
$\prn{f_!\,P}\prn{y} = \SigTy{x}{X}{\prn{f\,x = y} \times P\,x}$.
\looseness=-1 Opcartesian lifts along the injections and a vertical coproduct
assemble the sum-type clause of canonicity
(\cref{sec:gluing:unwinding}).

\subsubsection{Vertical Cartesian Closed Structure}

We bundle the vertical structures of this section into a single notion,
the input to the promotion theorems of \cref{sec:promotion}.

\begin{definition}[\AgdaRef{def:vertcc}]
  \label{def:vertcc}
  A displayed category $\disp{C}$ over a cartesian closed $\cat{C}$ is
  \emph{vertically cartesian closed} if it has vertical terminal objects
  (\cref{def:vertterm}), vertical binary products (\cref{def:vertbp}),
  cartesian lifts for all morphisms (i.e.\ $\disp{C}$ is a fibration,
  \cref{def:fibration}), vertical exponentials (\cref{def:vertexp}), and
  universal quantifiers for all displayed objects over products
  (\cref{def:uq}).
\end{definition}

The first four are familiar categorical
structures, lifted to the vertical setting: terminal objects, products,
and exponentials each yield a representing object in a single fiber
$\dispob{C}{c}$---though, as vertical universal properties, they also
characterize displayed morphisms from other fibers---and cartesian
lifts move objects between fibers. The last, the universal
quantifier, is qualitatively different: it relates objects across
fibers, as the right adjoint to a functor that weakens from
$\dispob{C}{x}$ to $\dispob{C}{x \times a}$. This cross-fiber operation is
needed to promote vertical exponentials to
\looseness=-1 displayed exponentials (\cref{thm:vert-cc}).

\begin{theorem}[\AgdaRef{thm:setd-vertical}]\label{thm:setd-vertical}
  \looseness=-1 $\SetD$ is vertically cartesian closed: it is a fibration
  (\cref{thm:setd-fibration}), and the remaining vertical universal
  properties are each computed pointwise over the elements of the base
  set:
  \begin{itemize}
    \item The vertical terminal object over $X$ sends each $x$ to the
      unit type.
    \item The vertical binary product of families $P$ and $Q$ sends
      each $x$ to $P\prn{x} \times Q\prn{x}$.
    \item The vertical exponential sends each $x$ to
      $P\prn{x} \to Q\prn{x}$.
    \item The universal quantifier $\uq{A}{P}$ sends $y$ to
      $\PiTy{a}{A}{P\prn{y, a}}$.
  \end{itemize}
\end{theorem}

\begin{intensionalremark}
  Typechecking these constructions against the universal properties
  for exponentials and quantifiers raises performance issues, addressed
  in \cref{sec:concrete:performance}.
\end{intensionalremark}

\subsection{Vertical-to-Displayed Promotion}
\label{sec:promotion}

A displayed model (\cref{def:dispmodel}) requires displayed
bicartesian closed structure over the syntactic category. Displayed
structure spans fibers, as a displayed product or exponential lies over
a different base object than the displayed objects it is built from.
Vertical structure is instead fiberwise, and over each
base object $\inob{c}{C}$ we require vertical universal properties.
In a fibration, cartesian lifts systematically promote vertical
structure to displayed structure. We treat the cartesian closed fragment. The
cocartesian fragment is dual (\cref{sec:duality}).

For products, cartesian lifts bring displayed objects into a common
fiber, where a vertical product can be taken. Exponentials require
an additional ingredient---the universal quantifier---to move
structure from a product fiber back to the exponential fiber.

\begin{theorem}[\AgdaRef{thm:vert-cart}]
  \label{thm:vert-cart}
  Suppose $\cat{C}$ is a cartesian category and $\disp{C}$ over
  $\cat{C}$ is a fibration with vertical terminal objects and vertical
  binary products. Then $\disp{C}$ is displayed cartesian.
\end{theorem}

\begin{proof}
  \clubpenalty=10000 Displayed terminal objects are given by vertical terminal objects,
  reindexed via cartesian lifts. Given $\mkdisp{x} : \dispob{C}{x}$ and
  $\mkdisp{y} : \dispob{C}{y}$, the displayed product over $x \times y$
  is the vertical product of their cartesian lifts along the
  projections,
  $\dispprod{\mkdisp{x}}{\mkdisp{y}} :=
  \vertprod{\cartlift{\pi_1}{\mkdisp{x}}}{\cartlift{\pi_2}{\mkdisp{y}}}$
  (\cref{fig:cartlift-prod}(b)).
\end{proof}
\clubpenalty=9999

This is the construction used to prove \cref{thm:canonicity}, where the
displayed product of the canonicity predicates for $\bool$ and $\nat$
was constructed as
$\vertprod{\cartlift{\pi_1}{\dispiota\prn{\bool}}}{\cartlift{\pi_2}{\dispiota\prn{\nat}}}$.

\begin{theorem}[\AgdaRef{thm:vert-cc}]
  \label{thm:vert-cc}
  Suppose $\cat{C}$ is cartesian closed and $\disp{C}$ over $\cat{C}$
  is vertically cartesian closed (\cref{def:vertcc}; in particular,
  $\disp{C}$ is a fibration). Then $\disp{C}$
  is displayed cartesian closed.
\end{theorem}

\begin{proof}
  The displayed terminal and product cases follow from
  \cref{thm:vert-cart}. Displayed exponentials require combining
  cartesian lifts, vertical exponentials, and the universal
  quantifier.

  Given $\mkdisp{a} : \dispob{C}{a}$ and
  $\mkdisp{b} : \dispob{C}{b}$, suppose $\cat{C}$ has an exponential
  $a \Rightarrow b$ with evaluation morphism
  $\app : \prn{a \Rightarrow b} \times a \to b$. The displayed
  exponential over $a \Rightarrow b$ is
  $\dispexp{\mkdisp{a}}{\mkdisp{b}} :=
  \uq{a}{\vertexp{\cartlift{\pi_2}{\mkdisp{a}}}{\cartlift{\app}{\mkdisp{b}}}}$.
  The construction proceeds in three steps:
  \begin{enumerate}
  \item cartesian lift $\mkdisp{a}$ along
    $\pi_2 : (a \Rightarrow b) \times a \to a$ and $\mkdisp{b}$ along
    $\app : (a \Rightarrow b) \times a \to b$, bringing both into the
    fiber over $(a \Rightarrow b) \times a$.
  \item Form the vertical exponential
    $\vertexp{\cartlift{\pi_2}{\mkdisp{a}}}{\cartlift{\app}{\mkdisp{b}}}$
    in the fiber over $(a \Rightarrow b) \times a$. This represents
    functions from $\mkdisp{a}$ to $\mkdisp{b}$ in the extended
    context.
  \item Apply the universal quantifier $\uq{a}{-}$ to move the
    exponential from the fiber over $(a \Rightarrow b) \times a$ down
    to the fiber over $a \Rightarrow b$, eliminating the dependency on
    the extra context $a$. \qedhere
  \end{enumerate}
\end{proof}

\begin{corollary}[\AgdaRef{cor:setd-dispcc}]\label{cor:setd-dispcc}
  $\SetD$ is displayed bicartesian closed.
\end{corollary}

\Cref{thm:setd-vertical} gives the cartesian closed fragment. Dually,
$\SetD$'s vertical initial objects, coproducts, and opcartesian lifts
are computed pointwise (\cref{sec:duality}). Together they produce
exactly the displayed bicartesian closed structure that a displayed
model requires.

\subsection{Assembling Displayed Models}
\label{sec:assemble}

A logical relation is given by a displayed model, which is built
using the elimination principle of a language's syntax
(\cref{thm:elim}). The eliminator asks for displayed structure matching
each connective of the language. Because we characterize each
connective semantically, independently of any specific syntax, this
structure can be assembled \emph{\`a la carte} from reusable semantic
components.

For each connective, we first
identify its universal property in the
syntactic category of the language. A displayed model must then
interpret the connective by the corresponding \emph{displayed}
universal property, and these displayed structures serve as the motive
of the eliminator for the syntax. Finally, we find \emph{vertical}
structures that are
(1) sufficient for constructing the
displayed structures via a \emph{vertical-to-displayed} theorem,
and
(2) stable under reindexing along a suitable class
of functors---namely, those preserving any universal
properties mentioned in the \emph{base category}.

\begin{construction}
  \label{constr:assemble}
  Let $\cat{S}$ be a free category over $\Sigma$ for a chosen set of
  connectives. To construct a displayed model over $\cat{S}$, it
  suffices to give
  \begin{enumerate}
    \item a semantic domain $\disp{C}$ over a base $\cat{C}$ with the
      vertical structures for each connective,
    \item a functor $F : \cat{S} \to \cat{C}$
      such that reindexing along $F$ preserves these vertical structures, and
    \item an interpretation
      $\dispiota : \interpIn{\Sigma}{\reindex{F}{\disp{C}}}$ of the
      generators.
  \end{enumerate}
  Reindexing along $F$ carries the vertical structure of $\disp{C}$
  to $\reindex{F}{\disp{C}}$, and the vertical-to-displayed theorem for
  each connective turns it into displayed structure over $\cat{S}$.
  The eliminator of $\cat{S}$ applied to $\dispiota$ then produces the
  displayed model.
\end{construction}

Recall the example language $\lang$ with unit, empty, product, sum, and
function types. To construct models of $\lang$ over a signature $\Sigma$,
we invoke \cref{constr:assemble} along a cartesian functor, with the
vertical and displayed bicartesian closed structures developed in this section.
The functor must be cartesian because the universal quantifiers used to
build displayed exponentials mention products in the base category.
The canonicity proof of \cref{sec:gluing} may
then be seen as this construction instantiated with $\SetD$ as the
semantic domain.

\section{Concrete Semantic Domains}\label{sec:concrete}

A proof in our framework begins by choosing a displayed category whose
displayed objects match the property being proved. One such displayed
category, families of sets $\SetD$, was developed alongside
the abstract theory of \cref{sec:dispuniv}. This section develops
three more. The displayed category of predicates is the
proof-irrelevant special case, suited to properties rather than
constructions. The displayed category of presheaves $\PshD$
generalizes $\SetD$ to logical relations that vary over contexts, as
in Kripke-style arguments, and is the basis of the conservativity
proof of \cref{sec:conservativity}. The isofiber and iso-comma
categories package isomorphisms rather than relations. They carry
displayed \emph{monoidal} structure and are the basis of the
coherence proof of \cref{sec:coherence}. We establish the vertical universal structures of $\PshD$---those
matching the type connectives in scope---once and for all, so the
promotion theorems of \cref{sec:promotion} apply to it as they did
for $\SetD$.
The isomorphism categories inherit their displayed structure
directly. The choice of concrete encoding has significant
consequences for typechecking performance, taken up in
\cref{sec:implementation}.

\subsection{The Displayed Category of Propositions}\label{sec:concrete:pred}

The canonicity relation of \cref{sec:gluing} is proof-relevant: for
each term it constructs an element of that term's set of canonical
forms. This type is a priori a \emph{set}. Since canonical forms are
unique, one could show after the fact that it is a \emph{proposition},
but the proof never relies on this.

Other metatheorems ask only to inhabit a \emph{proposition} at each
term, with no witnesses to extract. There the dependent eliminator
specializes to induction (\cref{sec:free-cats:elim}), and the building
block assigns propositions rather than sets: it is $\SetD$ valued in
homotopy propositions $\Prop$ rather than $\Set$.

\begin{definition}[\AgdaRef{def:predd}]\label{def:predd}
  The \emph{displayed category of propositions} $\PropD$ over $\Set$ has,
  over a set $X$, displayed objects the families $X \to \Prop$, and
  displayed morphisms over $f : X \to Y$ between families $P$ and $Q$ the
  fiberwise implications $\PiTy{x}{X}{P\prn{x} \to Q\prn{f\prn{x}}}$.
\end{definition}

Like $\SetD$, $\PropD$ is a vertically cartesian closed fibration.
Because the displayed data are propositions, the vertical universal
properties of $\PropD$ express closure conditions of propositions under each connective.

\subsection{The Displayed Category of Presheaves}\label{sec:concrete:psh}

For Kripke logical relations, the logical relation must vary over all
contexts rather than being defined at a single set of closed terms.
The displayed presheaf category provides this generality; the
conservativity proof of \cref{sec:conservativity} is the
instance developed in this paper.

\begin{definition}[\AgdaRef{def:pshd}]\label{def:pshd}
  Given a displayed category $\disp{C}$ over $\cat{C}$, the
  \emph{displayed presheaf category} $\PshD\prn{\disp{C}}$ over the
  presheaf category $\Psh\prn{\cat{C}}$ has, over a presheaf $P$,
  displayed objects the presheaves on
  $\totalcat{(\slicecat{\disp{C}}{P})}$, and displayed morphisms over a
  morphism of presheaves $\alpha : P \to Q$ the morphisms of displayed
  \looseness=-1 presheaves~(\cref{def:disppshmor}) over $\alpha$.
\end{definition}

\begin{theorem}[\AgdaRef{thm:pshd-vertcc}]\label{thm:pshd-vertcc}
  $\PshD\prn{\disp{C}}$ is a vertically cartesian closed fibration.
\end{theorem}
\cref{thm:vert-cc} therefore applies. The
rest of this subsection establishes the structure.

\paragraph{Fibration} The cartesian lift of a displayed presheaf
$\disppsh{Q}$ over $Q$ along a morphism of presheaves
$\alpha : P \to Q$ is precisely the reindexed presheaf
$\cartlift{\alpha}{\disppsh{Q}}$ of \cref{def:reindexdisppsh}.

\paragraph{Vertical structure} The vertical terminal objects and
products are computed pointwise in each fiber, mirroring the
constructions for $\SetD$ but at the presheaf level.

\subsubsection{Cocontinuity and Right Adjoints}

The difference from $\SetD$ is in the vertical exponentials and
universal quantifiers. In $\SetD$, these are computed pointwise:
the exponential of two families $P$ and $Q$ over $X$ sends each
$x$ to $P\prn{x} \to Q\prn{x}$. In $\PshD$, the analogous
constructions cannot be defined pointwise because the exponential of
two presheaves is not computed pointwise. Instead, the vertical
exponential and universal quantifier are each constructed by virtue
of being right adjoints.

We focus on the universal quantifier, which is the right adjoint to
the weakening functor that pulls displayed presheaves back along a
projection. Its existence rests on one principle:

\begin{theorem}[\AgdaRef{thm:cocont-adjoint}]
  \label{thm:cocont-adjoint}
  Every cocontinuous functor between presheaf categories has a right
adjoint.
\end{theorem}

This is an instance of the adjoint functor
theorem~\cite[Theorem~4.7.3]{riehl2017category}, and we give a
concrete account of it. In outline: by the coYoneda
lemma, every presheaf is canonically a colimit of representables, so
a cocontinuous functor $P : \Psh\prn{\cat{C}} \to \Psh\prn{\cat{D}}$
is determined by its action on representables, a profunctor
$R : \cat{D}^{\mathrm{op}} \times \cat{C} \to \Set$ with
$R\prn{d, c} = P\prn{\yo{c}}\prn{d}$. Rebuilding $P$ from $R$ is a
coend construction, the \emph{extension} of $R$, and the
extension of any profunctor has a right adjoint
$\larghom{R}{-}$, sending a presheaf $T$ to
$c \mapsto \Psh\prn{\cat{D}}\sqprn{R\prn{-, c}, T}$, the maps out of
the fibers of $R$. A cocontinuous $P$ is naturally isomorphic to the
extension of its own $R$, so it inherits this right adjoint. The full
\looseness=-1 development is in
\ifwithappendix\cref{app:cocont}\else the appendix of the supplementary material\fi.

The weakening functor of \cref{sec:uq} is a reindexing of presheaves
along the projection out of a slice category, and it sends
representables to representables. It is therefore cocontinuous, and
its right adjoint (\cref{thm:cocont-adjoint}) is the \emph{universal
quantifier} on $\PshD$.

\paragraph{The pushforward}
The right adjoints just constructed carry universal properties that
reach across fibers, and verifying them calls for one more piece of
presheaf infrastructure. Reindexing of displayed presheaves has a
left adjoint along any morphism of presheaves
$\alpha : P \to Q$, the \emph{pushforward} $\alpha_!$, defined
pointwise by a $\Sigma$-type: \(
  \prn{\alpha_!\,\disppsh{P}}\prn{\mkdisp{c}, q}
  \;=\;
  \SigTy{p}{\pshelt{P}{c}}{\disppshelt{P}{\mkdisp{c}}{p}
    \times \prn{\alpha\,p = q}}
\).
Read logically, the formula combines two connectives: it asserts
that \emph{there exists} an element $p$ satisfying $\disppsh{P}$
such that $\alpha\,p$ \emph{equals} $q$. The pushforward of
presheaves is thus built from the existential quantifier and the
equality type. Along a projection the equality contracts away and
the pushforward is existential quantification itself, the left
adjoint to the same weakening whose right adjoint is the universal
quantifier (\cref{sec:uq}); along a general $\alpha$, the equality
type accounts for the difference~\cite{lawvere1969adjointness}.

Maps of displayed presheaves $\alpha_!\,\disppsh{P} \to \disppsh{Q}$
over $Q$ correspond to maps
$\disppsh{P} \to \cartlift{\alpha}{\disppsh{Q}}$ over $P$; that is,
the pushforward is the opcartesian lift (\cref{sec:duality}) of a
displayed presheaf along $\alpha$, dual to the cartesian lift above.
A displayed morphism
over an arbitrary base map becomes a vertical morphism out of a
pushforward, so cross-fiber universal properties reduce to fiberwise
computations. Though the pushforward interprets no connective of
$\lang$, it underlies the right adjoints that do.
\emph{Frobenius reciprocity}~\cite[Lemma 1.9.12]{JacobsCLTT} says the existential
quantifier commutes with products, and
\emph{Beck--Chevalley}~\cite[Lemma 1.8.9]{JacobsCLTT} says it
commutes with substitution. We return to them in
\cref{sec:concrete:disppsh}.

\begin{intensionalremark}
  Verifying these universal properties in $\PshD$ is far more involved
  than for $\SetD$, and a naive encoding is unusably slow to typecheck
  (\cref{sec:concrete:pshmor}).
\end{intensionalremark}

\subsection{Displayed Categories of Isomorphisms}\label{sec:concrete:isofiber}

The isomorphism categories of this subsection package
\emph{isomorphisms} rather than relations. In the dependent-type
reading, the isomorphisms play the role of the equality type: a
section exhibits an isomorphism of a specified shape, turning
statements about inverses and natural isomorphisms into instances of
the elimination principle of \cref{sec:free-cats:elim}.

\begin{definition}[\AgdaRef{def:isoD}]\label{def:isoD}
  \looseness=-1 For a category $\cat{D}$, let $\Iso\prn{\cat{D}}$ be the displayed
  category over $\cat{D} \times \cat{D}$ whose displayed objects over a
  pair $\prn{d, d'}$ are the isomorphisms $\phi : d \cong d'$ in
  $\cat{D}$, and whose displayed morphisms over
  $\prn{g, g'} : \prn{d, d'} \to \prn{e, e'}$ from
  $\phi : d \cong d'$ to $\psi : e \cong e'$ are witnesses that $\seq{g}{\psi} = \seq{\phi}{g'}$.
\end{definition}

The isofiber and the iso-comma category are both reindexings of
$\Iso\prn{\cat{D}}$ along a functor into $\cat{D} \times \cat{D}$ that
chooses which two objects the isomorphism connects.

\begin{definition}[\AgdaRef{def:isofiber}]\label{def:isofiber}
  Given a functor $F : \cat{C} \to \cat{D}$, the \emph{isofiber}
  $\IsoFiber\prn{F}$ is the reindexing of $\Iso\prn{\cat{D}}$ along
  $\IdFun \times F : \cat{D} \times \cat{C} \to \cat{D} \times \cat{D}$,
  repackaged as a displayed category over $\cat{D}$ along the projection
  $\cat{D} \times \cat{C} \to \cat{D}$. Unfolding, its displayed objects
  over $\inob{d}{D}$ are pairs $\prn{c,\, \phi}$ with $\inob{c}{C}$ and
  $\phi : d \cong F\prn{c}$ an isomorphism in $\cat{D}$, and its
  displayed morphisms over $g : d \to d'$ between $\prn{c, \phi}$ and
  $\prn{c', \phi'}$ are morphisms $h : c \to c'$ in $\cat{C}$ such that
  $\seq{g}{\phi'} = \seq{\phi}{F\prn{h}}$.
\end{definition}

\begin{theorem}[\AgdaRef{thm:isofiber-reflection}]\label{thm:isofiber-reflection}
  A section of $\IsoFiber\prn{F}$ yields a functor $F^{-1} :
  \cat{D} \to \cat{C}$ such that
  $\seq{F^{-1}}{F} \cong \IdFun_{\cat{D}}$.
\end{theorem}

Beyond reasoning about isomorphisms, the isofiber carries algebraic
structure of its own. Unlike $\SetD$ and $\PshD$ it need not be
cartesian closed, but when $F$ is a strong monoidal functor it inherits
displayed monoidal structure, so the
eliminator for free monoidal categories targets it directly.
The coherence proof of \cref{sec:coherence} rests on this: a
section of it yields a retraction of $F$
(\cref{thm:isofiber-reflection}).

The iso-comma category is the same idea applied to two functors with a
common domain instead of one functor and the identity: a section will
exhibit a natural isomorphism between them.

\begin{definition}[\AgdaRef{def:isocomma}]\label{def:isocomma}
  Given functors $F, G : \cat{C} \to \cat{D}$, the \emph{iso-comma
    category} $\IsoComma\prn{F, G}$ is the reindexing of
  $\Iso\prn{\cat{D}}$ along
  $\langle F, G\rangle : \cat{C} \to \cat{D} \times \cat{D}$, displayed
  over the common domain $\cat{C}$. Its displayed objects
  over $\inob{c}{C}$ are isomorphisms $\phi : F\prn{c} \cong G\prn{c}$ in
  $\cat{D}$, and its displayed morphisms over $h : c \to c'$ between
  $\phi$ and $\phi'$ are proofs of naturality
  $\seq{F\prn{h}}{\phi'} = \seq{\phi}{G\prn{h}}$.
\end{definition}

\begin{theorem}[\AgdaRef{thm:isocomma-reflection}]\label{thm:isocomma-reflection}
  A section of $\IsoComma\prn{F, G}$ yields a natural isomorphism
  $F \cong G$.
\end{theorem}

When $\cat{C}$ is a free category and $\cat{D}$ carries the
corresponding structure, $\IsoComma\prn{F, G}$ inherits a displayed
structure of the same kind, so the eliminator targets it directly: a
section then witnesses that any two structure-preserving functors
agreeing on generators are naturally isomorphic. This is the mechanism
behind the uniqueness arguments in the conservativity proof
(\cref{sec:conservativity}).

\section{Applications}\label{sec:applications}

Each metatheorem in this paper instantiates \cref{constr:assemble}
with the appropriate set of connectives, using the eliminator of its
own free category (\cref{sec:free-cats:elim}): choose a semantic
domain, a functor from the syntax, and an interpretation of the
generators. When the semantic domain carries displayed structure
directly, as the isomorphism categories do
(\cref{sec:concrete:isofiber}), the vertical-to-displayed step is
skipped. In the canonicity proofs, including the step-indexed variant
of \cref{sec:stepindexed}, the section directly witnesses canonical
forms. For the remaining applications, a short argument after building
the section proves the property of interest.  Representation
\looseness=-1 independence combines the section with \emph{identity extension},
conservativity extracts \emph{full faithfulness}, and coherence
extracts \emph{thinness}.

The canonicity proof of \cref{sec:gluing} is the first instance of a
logical relation in this framework, and it is modular in the choice of
connectives. That is, we may perform the same proof of canonicity over
an arbitrary signature with any selection of universal
properties (terminals, initials, products, coproducts, exponentials)
and the negative fragment of the $\lambda$-calculus.
Across these canonicity theorems, the syntax may vary---and the choice of
universal properties is fit to the connectives in the syntax---but
the semantics is a reindexing of $\SetD$ each time.

The choice of semantic domain varies across applications.
In this section, parametricity uses
$\SetD$ (\cref{def:setd}), conservativity uses
$\PshD$ (\cref{def:pshd}), coherence uses $\IsoFiber$
(\cref{def:isofiber}), and step-indexed canonicity uses
$\FamwPsh$---a step-indexed variant of $\SetD$
(\cref{sec:stepindexed}).

\subsection{Parametricity}\label{sec:parametricity}
Reynolds' parametricity~\cite{reynolds1983types} captures the
idea that programs cannot depend on how an abstract type is
implemented. Parametricity is usually studied for languages with parametric
polymorphism, but the idea also applies to a simply typed language
parameterized by a signature, which we can think of as the language of an ML-style parameterized module.

Consider the signature $\SigFlip$ with a base type $\absty$ and
morphisms: $\opinit : \unitty \to \absty$,
$\opflip : \absty \to \absty$, and
$\opread : \absty \to \unitty + \unitty$. In $\Set$, we construct two
implementations of this interface. The
first, $\semB{\cdot}$, interprets $\absty$ as $\bool$, with
$\opinit = \boolt$, $\opflip = \notb$, and $\opread$ sending $\boolt$
to $\inl\;\unitel$ and $\boolf$ to $\inr\;\unitel$. The second,
$\semN{\cdot}$, interprets $\absty$ as $\nat$, with $\opinit = 0$,
$\opflip = \suc$, and $\opread = \even$, where $\even\;n = \inl\;\unitel$
exactly when $n$ is even.
We prove a representation independence
theorem~\cite{mitchell1986representation,powerrobinson2000structured},
showing that the two implementations are indistinguishable at types
that do not mention $\absty$.
Our development also proves a similar theorem for an interface to a stack of
booleans~(\AgdaRef{thm:parametricity-stack})
.

\begin{theorem}[Representation independence~\AgdaRef{thm:repind}]
  \label{thm:repind}
  For $\absty$-free types $A$ and $B$, the interpretation of every morphism
  $\incathomraw{e}{\BiCCC{\SigFlip}}{A}{B}$ is the same in each
  implementation, $\semB{e} = \semN{e}$.
\end{theorem}

To prove \cref{thm:repind}, we adapt the technique of Ghani
et~al.~\cite[Definition~4.1]{ghani2015bifibrational} for constructing
a displayed category suitable for \emph{binary} logical relations. The pairing
$\langle \semB{\cdot}, \semN{\cdot} \rangle : \BiCCC{\SigFlip} \to \Set \times \Set$
is itself a model (\cref{def:modellangsig}) because $\Set \times \Set$ is
bicartesian closed with its structure computed componentwise.
Composing with the product functor gives a cartesian functor
$F : A \mapsto \semB{A} \times \semN{A}$.

At $\absty$, a boolean and a natural are related when $\opread$ returns the
same value on both,
$\lrelat{\absty}{b}{n} := \prn{\semB{\opread}\,b = \semN{\opread}\,n}$;
that is, when $b$ is the parity of $n$. It is a short
proof to show that each of $\opinit$, $\opflip$, and $\opread$ respects
this relation.

The eliminator (\cref{thm:elim}) then extends the relation at $\absty$
to a section $R$ of
$\reindex{F}{\SetD}$, assigning each type $A$ a relation $\lrel{A}$ over
$\semB{A} \times \semN{A}$. Its action on morphisms is the
\emph{fundamental lemma} of the logical relation:
every morphism $\incathomraw{e}{\BiCCC{\SigFlip}}{A}{B}$ sends related
inputs to related outputs, so $\lrelat{A}{x}{y}$ implies
$\lrelat{B}{\semB{e}\;x}{\semN{e}\;y}$.

The fundamental lemma alone does not prove \cref{thm:repind}, because
it relates the outputs of the two implementations rather than
equating them. An identity extension lemma bridges the gap. Both
implementations interpret
$\absty$-free types identically, and on them the relation is equality.

\begin{theorem}[\AgdaRef{lem:identity-extension}]
  \label{thm:identity-extension}
  For every $\absty$-free $A$, $\semB{A} = \semN{A}$ and
  $\lrel{A}$ is the equality relation on it.
\end{theorem}
\begin{proof}[Proof of \cref{thm:repind}]
  Let $a \in \semB{A}$. $\lrel{A}$ is equality
  (\cref{thm:identity-extension}) and so $\lrelat{A}{a}{a}$ is inhabited.
  The fundamental lemma then relates $\semB{e}\;a$ and $\semN{e}\;a$ by
  $\lrel{B}$, which is again equality.
\end{proof}

\subsection{Conservativity}\label{sec:conservativity}
Adding function types to a language with products proves no new
equations between terms of the original language.
Categorically, the free
cartesian category $\CC{\Sigma}$ and the free cartesian closed
category $\CCC{\Sigma}$ are generated from the same signature, a
canonical functor $j$ interprets the former inside the latter, and
\emph{conservativity} is the statement that $j$ is fully faithful.

\begin{theorem}[\AgdaRef{thm:conservativity}]\label{thm:conservativity}
  Fix a signature $\Sigma$. The canonical functor
  $j : \CC{\Sigma} \hookrightarrow \CCC{\Sigma}$ embedding the free
  cartesian category into the free cartesian closed category is fully
  faithful.
\end{theorem}

Full faithfulness splits into two halves, which we treat separately.
\emph{Faithfulness} follows from freeness of $\CC{\Sigma}$ alone;
\emph{fullness} requires a logical relation built by gluing along the
nerve of $j$.

\subsubsection{Faithfulness}
$\Psh\,\CC{\Sigma}$ is a cartesian closed category, so an
interpretation $\kappa : \interpIn{\Sigma}{\Psh\,\CC{\Sigma}}$ induces
via the recursor a functor
$\recinterpCCC{\kappa} : \CCC{\Sigma} \to \Psh\,\CC{\Sigma}$. We take
$\kappa$ to send each generator of $\Sigma$, as interpreted in
$\CC{\Sigma}$, to its image under the Yoneda embedding.

Cartesian functors out of $\CC{\Sigma}$ that agree on
the generators are unique up to natural isomorphism. This is analogous
to the uniqueness of the recursor (\cref{sec:free-cats}), but up to
isomorphism instead of equality. Both $\yochar$ and
$\seq{j}{\recinterpCCC{\kappa}} : \CC{\Sigma} \to \Psh\,\CC{\Sigma}$
are cartesian and agree on the generators, so
$\yochar \cong \seq{j}{\recinterpCCC{\kappa}}$.

The Yoneda embedding is always faithful, and faithfulness is
invariant under natural isomorphism, so
$\seq{j}{\recinterpCCC{\kappa}} \cong \yochar$ is itself faithful. A
composite $\seq{j}{H}$ can be faithful only if its first factor $j$
is, so $j$ is faithful, completing half of the conservativity proof.

The uniqueness invoked here is itself an instance of
\cref{constr:assemble}, for the cartesian connectives and with an
isomorphism category as the chosen semantics. Since $\yochar$ and
$\seq{j}{\recinterpCCC{\kappa}}$ are both cartesian, the iso-comma
category $\IsoComma\prn{\yochar, \seq{j}{\recinterpCCC{\kappa}}}$
(\cref{def:isocomma}) carries displayed cartesian structure over
$\CC{\Sigma}$. The agreement of the two functors on generators gives
the displayed interpretation, and the cartesian eliminator produces a
section, which by \cref{thm:isocomma-reflection} is the natural
isomorphism $\yochar \cong \seq{j}{\recinterpCCC{\kappa}}$.

\subsubsection{Fullness}
Fullness demands a preimage for every morphism between objects in the
image of $j$, and the logical relation constructs these preimages. We build it on
$\PshD\,\CC{\Sigma}$, the displayed category of presheaves
(\cref{def:pshd}), reindexed along the nerve of $j$. This mechanizes
and generalizes Lafont's gluing argument for full faithfulness of the
canonical functor into a free cartesian closed category, which we
discuss in \cref{sec:discussion}. First, we define the functor that we
will glue along.

\begin{definition}
  \label{def:nerve}
  The \emph{nerve} of a functor $F : \cat{C} \to \cat{D}$ is the
  functor $\nerve{F} : \cat{D} \to \Psh\,\cat{C}$ sending an object
  $d$ to the presheaf $\yo{d}$ precomposed with $F$:
  $\nerve{F}\prn{d}\prn{c} = \cathom{D}{F\,c}{d}$.
\end{definition}

We then reindex $\PshD\,\CC{\Sigma}$ along $\nerve{j}$.
Because limits of presheaves are computed pointwise, the nerve of any
functor is cartesian. Therefore, reindexing along $\nerve{j}$
preserves the vertical cartesian closed structure of
$\PshD\,\CC{\Sigma}$, which we can then promote to displayed cartesian
closed structure.

We interpret $\Sigma$ in $\PshD\,\CC{\Sigma}$ by mapping base types to
displayed presheaves that track \emph{witnesses to fullness}. That is,
we define $\gamma : \interpIn{\Sigma}{\PshD\,\CC{\Sigma}}$ such that, for
a base type $A$, the displayed presheaf $\gamma\,A$ lies over the
presheaf $\nerve{j}\prn{A}$ and assigns to an object $B$ of
$\CC{\Sigma}$ and a morphism
$f : \cathomraw{\CCC{\Sigma}}{j\,B}{j\,A}$ the set of preimages of $f$
under $j$: \(
  \define
  {\prn{\gamma\,A}(B, f)}
  {\SigTy{g}{\cathomraw{\CC{\Sigma}}{B}{A}}{j\,g = f}}
\).
Using the eliminator, we obtain a section. With this
interpretation, however, the section does not immediately prove
fullness. Because
$A$ is required to be a base type, the section only proves fullness
\emph{into base types}: it supplies a preimage for every
$f : \cathomraw{\CCC{\Sigma}}{j\,B}{j\,A}$ whenever $A$ is a base
type.

Finally, we extend this to all types by performing a second
induction, this time over $\CC{\Sigma}$. We define a predicate over
$\CC{\Sigma}$ (\cref{sec:concrete:pred}): its displayed
object over a type $A$ is the proposition asserting fullness into $A$,
namely that every $f : \cathomraw{\CCC{\Sigma}}{j\,B}{j\,A}$ has a
preimage $g : \cathomraw{\CC{\Sigma}}{B}{A}$ with $j\,g = f$. The base
case is exactly the section above. The cartesian eliminator then
discharges the remaining cases: it combines witnesses for $A$ and $B$
into a witness for $A \times B$ and handles the terminal type. Since
the objects of $\CC{\Sigma}$ are generated from base types by the
terminal object and binary products, this exhausts all types. There
is no exponential case, because $\CC{\Sigma}$ has no function types. The
resulting global section produces a preimage for every morphism
\looseness=-1 between images of cartesian types, so $j$ is \emph{full}.

\subsection{Coherence for Monoidal Categories}\label{sec:coherence}

The coherence theorem for monoidal
categories~\cite{maclane1971categories} states that in the free
monoidal category $\Mon{X}$ on a set $X$, any two parallel
morphisms are equal: $\Mon{X}$ is \emph{thin}.

\subsubsection{The Displayed Category and Functor}

The proof instantiates \cref{constr:assemble} for the monoidal
connectives. The free category is $\Mon{X}$, a quotient-inductive type
with objects generated by $X$ and a binary tensor $\otimes$,
and morphisms generated by the monoidal natural isomorphisms
(associators and unitors) quotiented by the pentagon and triangle
equations. The semantic domain is $\IsoFiber\prn{F}$
(\cref{def:isofiber}) for a strong monoidal functor $F$ to be
defined below. It inherits a displayed
monoidal structure from $F$ directly, so the vertical-to-displayed
step is not needed. The functor $F$ comes from a discrete monoidal
category. The \emph{list category} $\List\prn{X}$ is the
discrete category of lists over $X$, with concatenation as the
monoidal product and the empty list as the unit. Being discrete,
$\List\prn{X}$ is in particular thin. The strong monoidal functor
$F : \List\prn{X} \to \Mon{X}$ sends a list to its
iterated tensor product.

\subsubsection{The Section and the Retraction}

To build the section we invoke the eliminator for $\Mon{X}$ and
interpret its generators in $\IsoFiber\prn{F}$. The
only generators are the elements $x \in X$. The
associators and unitors, together with the pentagon
and triangle equations relating them, are discharged automatically by
the displayed monoidal structure that $\IsoFiber\prn{F}$ inherits from
$F$. The interpretation therefore
reduces to a single choice: for each $x \in X$, a displayed object of
$\IsoFiber\prn{F}$ over $x$. By \cref{def:isofiber} such a displayed
object is a pair $\prn{c,\, \phi}$ of a list $c \in \List\prn{X}$ and an
isomorphism $\phi$ between $x$ and $F\prn{c}$ in $\Mon{X}$. We take the
singleton list $c = [x]$, whose image $F\prn{[x]} = x \otimes I$ is $x$
tensored with the unit, and for $\phi$ the right unitor
$\rho_x : F\prn{[x]} \cong x$. The eliminator extends this assignment
to a section of $\IsoFiber\prn{F}$ over $\Mon{X}$.
By \cref{thm:isofiber-reflection}, this section yields a functor
$F^{-1} : \Mon{X} \to \List\prn{X}$ together with a natural
isomorphism $\seq{F^{-1}}{F} \cong \IdFun_{\Mon{X}}$.

Conceptually, $F^{-1}$ sends each object of $\Mon{X}$ to a list of
generators---a fully right-nested tensor product---and the isomorphism
$\seq{F^{-1}}{F} \cong \IdFun_{\Mon{X}}$ is the coherence isomorphism
exhibiting every object as canonically isomorphic to such a list, just
as in Mac Lane's original argument. The strategy of extracting
coherence from a normalization function follows Beylin and
Dybjer~\cite{beylindybjer1996}.
$F^{-1}$ is only a \emph{one-sided} inverse to $F$. We never obtain
an equivalence, and the next step shows that thinness is exactly what
makes this one-sided retraction enough.

\subsubsection{Transferring Thinness}

Thinness transfers from $\List\prn{X}$ to $\Mon{X}$ along the retraction.

\begin{lemma}[\AgdaRef{lem:thin-isoretract}]\label{lem:thin-isoretract}
  Let $\cat{D}$ be thin. If $R : \cat{C} \to \cat{D}$ has a retraction
  $S$ with $\seq{R}{S} \cong \IdFun_{\cat{C}}$, then $\cat{C}$ is
  thin.
\end{lemma}
We apply the lemma with
$R = F^{-1}$ and $S = F$: the reflection theorem supplies the
isomorphism $\seq{F^{-1}}{F} \cong \IdFun_{\Mon{X}}$, exhibiting $F$ as
a retraction of $F^{-1}$. Since $\List\prn{X}$ is thin,
\cref{lem:thin-isoretract} applies:

\begin{theorem}[\AgdaRef{thm:coherence}]\label{thm:coherence}
  For any set $X$, the free monoidal category $\Mon{X}$ is thin.
\end{theorem}

Thinness is exactly Mac Lane's coherence theorem: any diagram of
structural isomorphisms in $\Mon{X}$ is a pair of
parallel morphisms, so thinness says every such diagram commutes.

\subsection{Step-Indexed Canonicity}\label{sec:stepindexed}

The canonicity relation of \cref{sec:gluing} does not extend to
languages with general recursion: a diverging term has no canonical
form. A standard remedy is a \emph{step-indexed} logical
relation~\cite{ahmed2006stepindexed}. In categorical terms such a relation is valued in
the \emph{topos of trees}~\cite{birkedal2012trees}, the category of
presheaves on the ordinal $\omega$. Recursion there is interpreted
with Nakano's later modality~\cite{nakano2000modality}: $\later A$ is
$A$ available one step later, and guarded recursive functions,
$\later A \to A$, have unique fixed points.

We follow the canonicity recipe of \cref{sec:gluing}, changing the
semantic domain from $\SetD$
to families in the topos of trees. This displayed category
generalizes $\SetD$: for any category $\cat{D}$, the displayed category $\Fam{\cat{D}}$
of $\cat{D}$-valued families over $\Set$ has vertical structure
inherited from $\cat{D}$, and $\SetD$ (\cref{def:setd}) is the case
$\cat{D} = \Set$. Taking $\cat{D}$ to be the topos of trees
\looseness=-1 yields $\FamwPsh$.

We isolate the structure needed to model guarded recursion as a
\emph{guarded logic}: a displayed category that is a fibration, has
vertical terminal objects, and carries a vertical later modality---a
vertical endofunctor $\vertlater$ with a vertical natural
transformation $\nextmap : \IdFun \Rightarrow \vertlater$, delaying an
$\omega$-presheaf by one step. The vertical structure of the family
construction is inherited from the category the families are drawn
from, so the vertical later modality comes directly from the later
modality of the topos of trees. Being vertical, a guarded logic is
preserved under reindexing.

The syntax is the free category on a signature with booleans, an
explicit delay $\delta$, and a guarded fixed-point operator
$\mathsf{fix}$ that supplies a fixed point for every guarded map
$\delta \star M$. Reindexing $\FamwPsh$ along the closed-terms functor
$\Cl$ glues the guarded logic onto this syntax, and the promotion
theorems of \cref{sec:promotion} promote its vertical structure for
the eliminator.

\begin{theorem}[Step-indexed
  canonicity~\AgdaRef{thm:stepindexed}]\label{thm:stepindexed}
  For every closed term $M$ of type $\bool$ and every fuel budget $n$,
  either $M = \delta^k\,b$ for some $k < n$ and $b \in \{\boolt,\boolf\}$
  ($M$ returns $b$ within $n$ steps), or $M = \delta^n\,N$ for some
  closed term $N$ ($M$ is still delayed after $n$ steps).
\end{theorem}

We carry out a second version of the same theorem
\emph{synthetically}~(\AgdaRef{synthetic}), in Guarded Cubical Agda, whose unsafe
\flag{guarded} and \flag{rewriting} flags add the later modality as a
language primitive. It reuses $\SetD$ and all
the guarded-logic modules of the analytic proof, so the only new work
is to establish a few properties of the vertical later modality as
constructed from the primitive later provided by Guarded Cubical Agda.
We take the analytic version as primary because it is safe.
The topos of trees, the later modality, and its fixed
points are all definable without extending the theory.

\section{Implementation in Intensional Type Theory}
\label{sec:implementation}

The mathematics of \crefrange{sec:dispuniv}{sec:applications} is independent
of any proof assistant. Realizing it in intensional type theory
forces decisions that never appear on paper: which equalities hold
definitionally, how universal properties are encoded, and how the
typechecker copes with those encodings. This section collects the
decisions. \Cref{sec:implementation:strictness} explains why the
vertical theory is a primitive notion rather than a special case of
the displayed one. \Cref{sec:concrete:performance} diagnoses the
transport chains that make naive instantiations intractable to
typecheck. \Cref{sec:implementation:eq,sec:concrete:pshmor} present
the remedy: an inductive-equality encoding in which the transports
compute away.

\subsection{Vertical Elements versus Displayed Elements}
\label{sec:implementation:strictness}

\widowpenalty=10000 Since $\id_c$ is the universal element of $\yo{c}$, one would hope
that a vertical universal element (\cref{def:vertunivelt}) is simply
a displayed universal element over $\id_c$
(\cref{def:dispunivelt}). In the Agda formalization it is not,
because of the composition action underlying
$\recvertyo{\mkdisp{e}}$: its value at a displayed morphism
$\mkdisp{f}$ lands in the fiber
over $\seq{f}{\id_c}$
rather than over $f$. The two fibers agree only up to the right unit
law $\seq{f}{\id_c} = f$, which in a general category is a
\emph{propositional}, rather than a \emph{definitional}, equality.
We therefore define the vertical universal element as a primitive
notion whose induced map lands over $f$ directly. We develop vertical
universal elements in parallel, providing a conversion to the displayed
universal element of identity. When the base category is strict---i.e.\
when the category laws hold definitionally---the two notions coincide.

\widowpenalty=9999

\subsection{The Transport Performance Problem}\label{sec:concrete:performance}

Verifying the vertical universal properties of $\SetD$ and $\PshD$
requires repeatedly realigning displayed data along the categorical
laws of the base. The source is the one seen in \cref{sec:dispuniv}: a
vertical morphism is a displayed morphism over the identity, so
combining vertical and displayed data yields a morphism over a
composite such as $\seq{f}{\id}$ or $\seq{\id}{\id}$, which coincides
with the intended morphism only up to a unit or associativity law---a
path $p$ rather than a definitional equality. Moving the data back over
the intended morphism is a \emph{transport}. Writing $\reindop\,p$ for
transport along $p$, vertical composition is the displayed composite
$\reindop$-ed along the unit law $\seq{\id}{\id} = \id$.

When several universal properties are composed---as in the displayed
products, exponentials, and quantifiers of \cref{sec:dispuniv}, built
from projections, application, and weakening---these transports stack
into chains: \(
  \reindop\,p_n\big(\cdots\reindop\,p_2\big(\reindop\,p_1\,(x)\big)\big)
\).
Such goals are discharged by a \emph{chain of fillers}, peeling back
one transport at a time along the law responsible for it. Fillers are
a common Cubical Agda idiom implementing the abstract
constructions of \cref{sec:dispuniv}.

The difficulty is specific to concretely instantiating these
constructions for $\SetD$ and $\PshD$. The chain-of-fillers strategy
often produces a proof term of reasonable size and correct type, yet
sometimes it is prohibitively slow for Agda to typecheck. We hypothesize
that term size is a major contributor. Each layer of $\reindop$
appears to carry large implicit arguments that later layers grow and
duplicate, so the term size appears to grow with the depth of the
chain.

The severity scales with the depth of the chains. For $\SetD$ this
remains tractable: the cartesian structure (terminal objects, binary
products, fibration) typechecks with acceptable performance, and
although the deeper obligations for exponentials and universal
quantifiers are orders of magnitude slower, they still go through. For
$\PshD$, whose connectives stack still more transports, we have never
seen typechecking of the exponential and quantifier obligations
terminate, even with flags such as \flag{lossy-unification} that
accelerate unification with incomplete heuristics.

For both $\SetD$ and $\PshD$, most accumulated transports are over
reflexivity. Transporting over refl is propositionally but not
definitionally the identity function: each such transport remains a
stuck term that the elaborator must carry through unification rather
than simplify away. These stuck terms complicate an intuitive
description of our proofs, and, worse, they vastly slow down the speed
of typechecking. To address this, we adopt an equality for which
transporting over refl \emph{does} reduce.

\subsection{Inductive Equality}
\label{sec:implementation:eq}

The solution is to replace path equality with Martin-L\"{o}f's
inductive equality (\texttt{Eq}) in the infrastructure that mediates
universal properties. Transporting over refl with \texttt{Eq}'s $J$
eliminator is \emph{definitionally} the identity function, so the
accumulated transports compute away and the bottleneck disappears.
We only use \texttt{Eq} to establish universal properties in specific
categories such as $\SetD$ and $\PshD$. We then convert these proofs
back to the path-based universal properties of \cref{sec:dispuniv}, which the rest
of the library uses. The development of
\cref{sec:dispuniv} itself needs none of this machinery.

\subsection{Strict Presheaf Morphisms}\label{sec:concrete:pshmor}

The \texttt{Eq}-based encoding demands that the base category be
strictly associative and unital: the categorical laws must hold
definitionally.
For $\Set$, this is immediate; however, the standard definition
of natural transformations between presheaves yields a presheaf
category $\Psh\prn{\cat{C}}$ whose identity and associativity laws
hold only \emph{propositionally}.

We instead provide a definition of morphism between presheaves that
is definitionally unital and associative.
The strict encoding is obtained from the standard one by two
observations: $\eta$-equality for record types, which lets morphisms be
compared componentwise, and
a \emph{Yoneda encoding} of equations,\footnote{This is an instance of the Yoneda
  lemma for identity types.} which replaces a
term with a fresh variable constrained by an equation.
This encoding resembles the ``fording'' trick for
representing inductive families from McBride's
thesis~\cite{mcbride1999thesis}.
Kaposi and Pujet's strictified
syntax~\cite{kaposi2025strictification} made clear to us that combining
such encodings with $\eta$-equality yields additional definitional
equalities for record types of this kind.

Concretely, a morphism of presheaves in either encoding has
components $\alpha_c : \pshelt{P}{c} \to \pshelt{Q}{c}$; the
encodings differ in how naturality is stated. Without the Yoneda encoding,
naturality is an equation about a composite: for
$f : \cathom{C}{c}{c'}$ and $p' \in \pshelt{P}{c'}$, \(
  \alpha_c\prn{\pshelt{P}{f}\prn{p'}}
  = \pshelt{Q}{f}\prn{\alpha_{c'}\prn{p'}}
\).
The action of $P$ sits inside the argument of $\alpha$, so composing
two naturality proofs requires path algebra---congruence and
concatenation---which is associative and unital only up to further
paths. The encoding introduces a fresh variable
$p$ constrained by an equation and states naturality as a function
between equations: \(
  \pshelt{P}{f}\prn{p'} = p
  \;\longrightarrow\;
  \pshelt{Q}{f}\prn{\alpha_{c'}\prn{p'}} = \alpha_c\prn{p}
\).
Now the identity morphism's naturality proof is the identity
function, composing naturality proofs is composing functions, and
function composition is strictly associative and unital. Together
with $\eta$-equality, the category
laws for strict morphisms compute strictly.
The two encodings are isomorphic, but only the strict one
is suitable for the universal property proofs of $\PshD$ to typecheck
fast enough to use in practice.

\subsection{Displayed Presheaves}\label{sec:concrete:disppsh}

We encode a displayed presheaf as
an ordinary presheaf on a total category (\cref{def:displayedpsh}).
We prove once
and for all the two coherence laws that the pushforward of
\cref{sec:concrete:psh} must satisfy to behave like an
existential quantifier~\cite{JacobsCLTT}.
\emph{Frobenius reciprocity} is the
categorical form of the law that $\exists x.\,\prn{\phi \wedge \psi}$
agrees with $\prn{\exists x.\,\phi} \wedge \psi$ when $\psi$ does not
mention $x$. \emph{Beck--Chevalley} is the categorical form of the law
that existential quantification commutes with substitution, so that
$\prn{\exists x.\,\phi}\sqprn{t/y}$ agrees with
$\exists x.\,\prn{\phi\sqprn{t/y}}$.
In the standard fibrational account of quantifiers, these two laws are
proof obligations that a user must verify separately.

In our setting, for a morphism of presheaves $\alpha : P \to Q$ and
displayed presheaves $\disppsh{P}$ over $P$ and $\disppsh{Q}$ over $Q$,
Frobenius reciprocity is given by the vertical isomorphism of
displayed presheaves
\(
  \alpha_!\prn{\vertprod{\disppsh{P}}{\cartlift{\alpha}{\disppsh{Q}}}}
  \cong
  \vertprod{\alpha_!\,\disppsh{P}}{\disppsh{Q}}
\)~(\AgdaRef{frobenius}).
Because a displayed presheaf is a presheaf, it suffices to show a
natural family of pointwise isomorphisms in $\Set$.
At a point $q$ of $Q$, both
sides unfold to $\Sigma$-types over elements $p$ of $P$ with
$\alpha\,p = q$. On the left the $\disppsh{Q}$ component lies over
$\alpha\,p$, and on the right it lies over $q$. The isomorphism
reassociates the $\Sigma$-types and transports the $\disppsh{Q}$
component along $\alpha\,p = q$.

For Beck--Chevalley, let $\alpha : P \to R$, $\beta : Q \to R$,
$\alpha' : S \to Q$, and $\beta' : S \to P$ be morphisms of presheaves
such that $\seq{\alpha'}{\beta} = \seq{\beta'}{\alpha}$ and this
square is a pullback.
For a displayed presheaf $\disppsh{P}$ over $P$, Beck--Chevalley is
given by the vertical isomorphism of displayed presheaves
\(
  \alpha'_!\prn{\cartlift{\beta'}{\disppsh{P}}}
  \cong
  \cartlift{\beta}{\prn{\alpha_!\,\disppsh{P}}}
\)~(\AgdaRef{beck-chevalley}).
We again show it holds pointwise in $\Set$. At a point $q$ of $Q$, the left side
unfolds to a $\Sigma$-type over elements $s$ with $\alpha'\,s = q$,
and the right side to a $\Sigma$-type over elements $p$ with
$\alpha\,p = \beta\,q$. Because the square is a pullback, the two
$\Sigma$-types are in bijection.

These two laws hold in the displayed category of displayed
presheaves. Because the universal properties in our framework are
characterized via representability conditions on displayed
presheaves, the displayed Yoneda lemma lets us establish these laws
once for $\PshD$ when deriving the promotion theorems for any displayed
category~(\cref{sec:promotion}). Users of our framework never prove
them as side conditions.

\section{Related and Future Work}
\label{sec:discussion}

\subsection{Related Work}

\paragraph{Fibered category theory}\label{sec:fibered}
The formalisms throughout \cref{sec:dispuniv} adapt well-studied
notions from the theory of fibrations~\cite{JacobsCLTT}.
Conventionally, a \emph{fibration} over $\cat{C}$ is a functor
$p : \cat{E} \to \cat{C}$ such that there exists a cartesian lift of
every object along morphisms of $\cat{C}$.
A displayed category
corresponds to the projection out of its total
category~\cite{ahrens2019displayed}. \cref{def:fibration}, which
chooses a cartesian lift for each object and morphism, makes this
projection a \emph{cloven} fibration.
When $\cat{E}$ and $\cat{C}$ are univalent, the mere existence of
cartesian lifts determines a unique cleavage. Because chosen lifts have
good definitional behavior, we take displayed categories with chosen
lifts as our data structure
(\cref{sec:implementation}).

Fibrations are equivalent to \emph{fibered
categories}, pseudofunctors from $\op{C}$ into the bicategory of
categories.
The pseudofunctor assigns to each base morphism
$\incathom{f}{C}{d}{c}$ a \emph{reindexing functor}
$\cartliftbare{f} : \fibercat{C}{c} \to \fibercat{C}{d}$ between
fiber categories.
Further, displayed categories over $\cat{C}$ are extensionally equivalent to
normal lax double functors from $\cat{C}$ into a double category of
profunctors~\cite{benabou2000distributors}. An object $c$ is mapped to
the fiber $\fibercat{C}{c}$,
and a morphism $\incathom{f}{C}{d}{c}$ is mapped to the profunctor of
displayed morphisms
$\disphom{C}{f}{-}{-} : \fibercat{C}{d} \nrightarrow
\fibercat{C}{c}$. Even though reindexing \emph{functors} only exist
for fibrations, reindexing \emph{profunctors} over any base morphism
exist in any displayed category. A fibration is the case where each
of these profunctors is represented by $\cartliftbare{f}$:
\(
  \disphom{C}{f}{\mkdisp{d}}{\mkdisp{c}} \;\cong\;
  \cathomraw{\fibercat{C}{d}}{\mkdisp{d}}{\cartlift{f}{\mkdisp{c}}}
\).

A fibered category exhibits a \emph{fibered universal
  property} when each fiber category has a universal property such
that all reindexing functors preserve it~\cite{JacobsCLTT}.
In our setting, a \emph{vertical} universal property over $c$ is the
same data as a universal property in the fiber over $c$ such that each
reindexing profunctor preserves it. Since reindexing profunctors
always exist, vertical universal properties make sense even when
$\disp{C}$ is not a fibration.
When $\disp{C}$ is a fibration, each reindexing profunctor is
represented by $\cartliftbare{f}$, and vertical universal properties
coincide with fibered universal properties.

Modulo the coincidence of vertical and fibered universal
properties, the promotion theorems of \cref{sec:promotion} correspond
to the work of Hermida~\cite[Corollary~3.3.10]{hermida1993fibrations}. He shows that fibered
cartesian closed structure, along with universal quantifiers, induces
a cartesian closed structure on the total category that is preserved
by the projection. A cartesian closed total category whose projection
preserves that structure is equivalent to a displayed cartesian closed
category, so \cref{thm:vert-cc} is
analogous to Hermida's corollary.
In his setting the quantifiers must satisfy the
Beck--Chevalley condition~\cite[Definition~1.4.4]{hermida1993fibrations},
but in our encoding it is a theorem (\cref{sec:concrete:disppsh}).

\paragraph{Type theory in type theory}
The metatheory of type theory is increasingly formalized using an
\emph{intrinsic}, quotiented syntax, in which well-typed terms are
identified up to definitional equality from the outset. Kaposi and
Pujet~\cite{kaposi2025strictification} strictify the substitution
calculus so that its coherence equations hold definitionally, and
Chen, Nordvall Forsberg, and Tsai~\cite{chen2026formalise} give a
comparable development in Cubical Agda; both model the object language
as a category-with-families with dependent products and booleans. Our
object languages are simply typed rather than dependent, and
metatheorems follow from the displayed-category eliminator, but we
share the concern with coherence of intrinsic syntax. A strictified
syntax obviates coherence issues; our approach does not commit to
strictness a priori, yet even when strictness is not required it
improves performance in practice.

\paragraph{Gluing as a metatheoretic technique}

The fibrational view of logical predicates originates in Hermida's
thesis~\cite{hermida1993fibrations}; our canonicity proof is a
displayed presentation of his gluing construction (\cref{sec:gluing}).
Lafont~\cite[Annexe C]{lafont1988canonical} proves that the canonical
functor from a category into the free cartesian closed category on it
is full and faithful, arguing fullness by gluing along the nerve into
presheaves so that the glued category inherits cartesian closed
structure. We prove a family of such conservativity theorems: besides
\cref{thm:conservativity}, the canonical functors from the free
category on $\Sigma$ into $\CC{\Sigma}$ and into $\CCC{\Sigma}$ are
fully faithful, each by the same argument instantiated for different
connectives. The theorem for the free category on $\Sigma$ into
$\CCC{\Sigma}$ is closest to Lafont's, taking a signature rather than a
category. Where Lafont builds the glued category and checks its
cartesian closed structure by hand, we reindex the displayed category
of presheaves along the nerve and obtain that structure from
\cref{constr:assemble}.

Gluing adapts to substructural settings as well.
Hasegawa~\cite{hasegawa1999logical} studies logical predicates for
intuitionistic linear type theory, a monoidal setting adjacent to our
coherence theorem (\cref{sec:coherence}).
Shulman~\cite{shulman2021envelopes} proves the universal map from a
closed symmetric monoidal category into its free \(*\)-autonomous
completion fully faithful, deducing this from the double gluing of
Hyland and Schalk~\cite{hylandschalk2003} with a technique of
Lafont---which our proof mechanizes in the cartesian
setting.

For dependent type theories, gluing has been progressively
internalized: Kaposi, Huber, and Sattler~\cite{kaposi2019gluing} for
canonicity and normalization, Sterling~\cite{sterling2021thesis}
through synthetic Tait computability, and Bocquet, Kaposi, and
Sattler~\cite{bocquet2023internal} via internal sconing. These work
internally to a chosen model or through a bespoke glued category. Our
constructions are analytic and generic in the model, packaging
gluing as reusable semantic components that apply across
languages.

\paragraph{Fiberwise structure and representability}
Blanco and Zeilberger~\cite{blancozeilberger2020bifibrations} show
that the multiplicative connectives of linear logic admit two agreeing
descriptions: through universal properties (a representable
polycategory) and through fibrational ones (a bifibration over the
terminal polycategory). The same phenomenon appears here: we formalize
that a cartesian closed category is equivalent to a vertically
cartesian closed category displayed over the terminal category
$\termcat$. This underlies the recursor special case of the eliminator
(\cref{sec:free-cats:elim}), where the displayed data are independent
of the syntax, and the promotion of \cref{sec:promotion} generalizes
the correspondence.

\paragraph{Mechanized logical relations}
A prominent application for mechanized logical relations is the
metatheory of Martin-L\"of type theory. Abel et
al.~\cite{abel2018decidability} formalize a conversion checker and
prove it correct with an inductive-recursive Kripke logical relation. Adjedj et
al.~\cite{adjedj2024martinlof} adapt this approach to formalization in
Coq (Rocq); moreover, they remove the reliance on induction-recursion and
additionally prove decidability of typechecking.
Poiret et al.~\cite{poiret2026divide} formalize in Rocq a
proof-irrelevant logical relation based on partial equivalence
relations (PERs). They write that categorical abstractions are
powerful, yet they ``remain challenging to reconcile with concrete mechanized
developments.'' Our work targets the tension between the
expressivity and concision of powerful abstractions and the pragmatic
concerns of proof engineering. These works mechanize logical relations
for a particular theory, while our framework aims to give a library of
abstractions that may be reused across theories. Our abstractions so far
compositionally build logical relations for simply typed theories, but
do not yet handle dependent types as these works do.

\paragraph{Mechanized category theory}
Several libraries formalize category theory in proof assistants:
agda-categories~\cite{hu2021formalizing}, UniMath~\cite{UniMath}, the
\texttt{agda/cubical} standard library~\cite{cubicalagdalib}, and
1Lab~\cite{1lab}. agda-categories differs from the rest in representing
equality of morphisms with setoids---an equivalence relation carried
alongside the morphisms. The remaining libraries, ours included, use
\emph{paths}: an equality type without uniqueness of identity proofs,
as in homotopy type theory. Among them, UniMath, in Rocq, is the
original home of displayed categories~\cite{ahrens2019displayed}, but
there univalence and function extensionality are postulated as axioms
and thus have no computational content, so terms that transport along
them get stuck. \texttt{agda/cubical}, which we build atop, and 1Lab
are built in Cubical Agda, where these instead compute. We differ from
all of them in motivation, as our development is oriented toward
compositional semantic constructions of logical relations rather than
general-purpose category theory. For instance, these libraries provide
separate records for each universal property, whereas we treat them
uniformly, as universal elements of (displayed) presheaves.

\subsection{Future Work}

Our signatures have generating types and terms but no equational
axioms, so the categories they generate are \emph{free}. Many theories
of interest instead impose equations---such as those of an inductive
recursor---and handling them is a natural next step.

We aim to apply the techniques of this paper to
languages with effects, substructural types, and dependent types.
Effectful languages in call-by-push-value~\cite{levy2003cbpv} have
categorical semantics that fit our framework, and our step-indexed
canonicity theorem (\cref{sec:stepindexed}) is a first step
toward logical relations for languages with richer effects.
For substructural languages, the monoidal Grothendieck
construction~\cite{moeller2020monoidal} already supplies a monoidal
analogue of our vertical-to-displayed promotion, and results like
Hasegawa's linear logical predicates~\cite{hasegawa1999logical} and
Shulman's $*$-autonomous conservativity~\cite{shulman2021envelopes} are
the kind of metatheorem we would target. The semantic layer
makes no simple-typing assumptions, and the remaining work for
dependent type theory lies in finding a suitable presentation of the
dependent syntax and its elimination principle.

\begin{acks}
  We thank Andreas Nuyts and Josselin Poiret for helpful discussions,
  the developers of 1Lab and the maintainers of the \texttt{agda/cubical}
  library, and the anonymous reviewers for their feedback. We also
  thank Eric Bond, Pranav Srinivasan, Cass Alexandru, Nathan Varner, and
  Ivan Wei for their contributions to the library.

  This material is based upon work supported by the Air Force Office of
  Scientific Research under Grant No.~\grantnum{}{FA9550-23-1-0760} and the U.S. National Science Foundation CAREER Award \grantnum{}{2540652}. Any
  opinions, findings, and conclusions or recommendations expressed in this
  material are those of the authors and do not necessarily reflect the views of
  the U.S. Department of Defense or the U.S. National Science Foundation.
\end{acks}

\section*{Disclosure of Use of AI in Research}
We used the AI coding assistant Claude Code (Anthropic) in developing
the Agda formalization, in the following ways:
\begin{itemize}
  \item \emph{Routine code.} We used AI to derive the implementation of
    coproducts, and of their displayed counterparts, from the existing
    implementation of products by duality, which was a mechanical port. We also
    used AI to assemble facts about cartesian functors for
    the parametricity proof of \cref{sec:parametricity}, such as
    closure under composition and pairing.
  \item \emph{Examples.} After we constructed the section establishing
    canonicity, we used AI to write examples that apply it.
    One example is an isomorphism between the closed terms of $\bool + \nat$ and the
    sum type of booleans and naturals in Agda.
    These only serve as a sanity check that the theorem is usable, and
    no result in the paper depends on them.
  \item \emph{Development tooling.} We used AI to aid with git
    operations and for preliminary code review.
\end{itemize}
All definitions, interfaces, and proof strategies were designed by the
authors. All AI-assisted code has been reviewed and endorsed by the
authors.

\section*{Data Availability Statement}
The Cubical Agda library and formalization of our proofs are
archived on Zenodo~\cite{schaefer2026artifact}, and the source code is
available on
\href{https://github.com/um-catlab/cubical-categorical-logic}{GitHub}.
The artifact typechecks with versions of Agda and the
\texttt{agda/cubical} library pinned in a Nix flake. All results
are checked in the \flag{safe} fragment of Agda except the synthetic
development of step-indexed canonicity (\cref{sec:stepindexed}), which
requires the guarded extension of Cubical Agda.

\bibliography{refs.bib}

\ifwithappendix
\appendix
\crefalias{section}{appendix}

\section{The Concrete Account of Cocontinuity}
\label{app:cocont}

This appendix expands the outline behind \cref{thm:cocont-adjoint}:
the tensor of presheaves, the coYoneda lemma, extension, the precise
definition of cocontinuity, and the adjunction.

\paragraph{The tensor of presheaves}
Recall that a morphism $\incathom{f}{C}{x}{c}$ acts on a point
$q \in \pshelt{Q}{c}$ to give a point $\pshelt{Q}{f}\prn{q}$ of
$\pshelt{Q}{x}$, a kind of composition of $f$ with $q$
(\cref{sec:univelts}).
The tensor performs this composition for a whole covariant family of
inputs at once.
\begin{definition}[Tensor of presheaves]
  \label{def:psh-tensor}
  Given a copresheaf $P : \cat{C} \to \Set$ and a presheaf
  $Q : \cat{C}^{\mathrm{op}} \to \Set$, their \emph{tensor}
  $\pshtens{P}{Q}$ is the coend $\int^{c} P\,c \times \pshelt{Q}{c}$: the
  set of formal composites $\pshtens{s}{q}$ of an input $s \in P\,c$ with
  a point $q \in \pshelt{Q}{c}$, for any object $c$, identified by the
  rule that a morphism may slide from the input to the point,
  \[
    \pshtens{\prn{P\,g\,s}}{q}
    \;=\;
    \pshtens{s}{\prn{\pshelt{Q}{g}\prn{q}}}
    \qquad
    \prn{\incathom{g}{C}{c}{c'},\; s \in P\,c,\; q \in \pshelt{Q}{c'}}.
  \]
\end{definition}
The coend identifies the two ways of composing: pushing the
covariant data forward, $P\,g : P\,c \to P\,c'$, and pulling the
contravariant data back,
$\pshelt{Q}{g} : \pshelt{Q}{c'} \to \pshelt{Q}{c}$, yield the same
formal composite. When the copresheaf $P$ varies functorially in a
further parameter, $\pshtens{P}{Q}$ becomes a presheaf in that
parameter.

When the copresheaf is a representable, the tensor reconstructs the
presheaf it acts on:
\begin{theorem}[coYoneda lemma]
  \label{thm:coyoneda}
  For every presheaf $Q : \cat{C}^{\mathrm{op}} \to \Set$ and object
  $\inob{x}{C}$, the map
  $\pshelt{Q}{x} \to \pshtens{\cathom{C}{x}{-}}{Q}$ sending $q$ to
  $\pshtens{\id}{q}$ is an isomorphism, natural in $x$; its inverse sends
  $\pshtens{f}{q}$ to $\pshelt{Q}{f}\prn{q}$.
\end{theorem}

\paragraph{Extension}
The coYoneda lemma reconstructs a presheaf by tensoring against the
representable copresheaves $\cathom{C}{x}{-}$ of its own category.
Tensoring against copresheaves from a different category
instead transports the presheaf across categories, turning a
profunctor into a functor on presheaves.
\begin{definition}[Extension]
  \label{def:extension}
  A functor $R : \cat{D}^{\mathrm{op}} \times \cat{C} \to \Set$ extends
  to a functor $\pext{R} : \Psh\prn{\cat{C}} \to \Psh\prn{\cat{D}}$,
  acting on a presheaf $Q$ by
  \[
    \pext{R}\prn{Q} \;:=\; \prn{\inob{d}{D} \mapsto \pshtens{R\,d}{Q}}
  \]
  where the copresheaf $R\,d$ is obtained by fixing the first argument
  of $R$.
\end{definition}
Taking $R = \cathom{C}{-}{-}$ recovers the identity functor: by
\cref{thm:coyoneda}, $\pext{\cathom{C}{-}{-}}\prn{Q} \cong Q$.

\paragraph{Cocontinuity}
A functor between cocomplete categories is cocontinuous when it
preserves colimits. Between presheaf categories this takes a concrete
form: since every presheaf is a colimit of representables
(\cref{thm:coyoneda}), a colimit-preserving functor is pinned down by its
values on representables, and rebuilding it from those values is an
extension.
\begin{definition}[Cocontinuity]
  \label{def:cocont}
  Let $P : \Psh\prn{\cat{C}} \to \Psh\prn{\cat{D}}$ be a functor and
  let $R : \cat{D}^{\mathrm{op}} \times \cat{C} \to \Set$ be its
  action on representables,
  $R\prn{d, c} := \prn{P\prn{\cathom{C}{c}{-}}}\,d$. Then $P$ is
  \emph{cocontinuous} when, for every presheaf
  $Q : \cat{C}^{\mathrm{op}} \to \Set$, the canonical comparison map
  $\pext{R}\prn{Q} \to P\prn{Q}$ is an isomorphism.
\end{definition}
\paragraph{Right adjoints from cocontinuity}
Cocontinuity is what guarantees a right adjoint, because an extension
always has one. The extension $\pext{R}$ is left adjoint to the functor
$\prn{\larghom{R}{-}} : \Psh\prn{\cat{D}} \to \Psh\prn{\cat{C}}$ that
sends a presheaf $T$ to
$c \mapsto \Psh\prn{\cat{D}}\sqprn{R\prn{-, c}, T}$, the maps out of
$R\prn{-, c}$.
\begin{theorem}
  \label{thm:adjoint}
  $\pext{R} \dashv \prn{\larghom{R}{-}}$: maps
  $\pext{R}\prn{S} \to T$ correspond naturally to maps
  $S \to \larghom{R}{T}$, for every presheaf $S$ on $\cat{C}$ and $T$
  on $\cat{D}$.
\end{theorem}
A cocontinuous $P$ is naturally isomorphic to an extension, so it
inherits this right adjoint: combining \cref{def:cocont} with
\cref{thm:adjoint},
\[
  \Psh\prn{\cat{D}}\sqprn{P\prn{S}, T}
  \;\cong\;
  \Psh\prn{\cat{D}}\sqprn{\pext{R}\prn{S}, T}
  \;\cong\;
  \Psh\prn{\cat{C}}\sqprn{S, \larghom{R}{T}},
\]
so $\prn{\larghom{R}{-}}$ is right adjoint to $P$.

\fi %

\end{document}